\documentclass[useAMS, usenatbib, a4paper]{mn2e}

\usepackage{mathrsfs}
\usepackage[space]{grffile}
\usepackage[T1]{fontenc}
\usepackage{aecompl}
\usepackage{times}
\usepackage{subfigure}
\usepackage{amsmath, amssymb}
\usepackage{amsfonts}
\usepackage{graphicx}
\usepackage{multirow}
\usepackage[T1,hyphens]{url}
\usepackage{bm}
\usepackage{ulem} %remove once not using sout
\usepackage[abs]{overpic}
\usepackage{tikz}
\usepackage{todonotes}
\usepackage{times}
\usepackage{amssymb}
\usepackage{amsmath}
\usepackage{graphicx,color}
\usepackage{bmpsize}
\usepackage{epstopdf}
\usepackage{adjustbox}
\usepackage{multicol}
\usepackage{rotate}

\usepackage{hyperref}

\newcommand{\vect}[1]{\mbox{\boldmath ${#1}$}}
\newcommand {\apgt} {\ {\raise-.5ex\hbox{$\buildrel>\over\sim$}}\ }
\newcommand {\aplt} {\ {\raise-.5ex\hbox{$\buildrel<\over\sim$}}\ }

\newenvironment{packed_enum}{
\begin{enumerate}
  \setlength{\itemsep}{1pt}
  \setlength{\parskip}{0pt}
  \setlength{\parsep}{0pt}
}{\end{enumerate}}

\def\myputfigure#1#2#3#4#5%
{\vskip#5pt\makebox[0pt]{\hskip#2in
\includegraphics[width=#3\textwidth]{#1}}\vskip#4pt\hfill}

\newcommand{\cmfast}{\textsc{\small 21CMFAST }}

\newcommand{\oskar}{\textsc{\small OSKAR }}
\newcommand{\wsclean}{WS C\textsc{\small LEAN }}

\newcommand{\invmpc}{cMpc$^{-1}$}
\newcommand{\klos}{k^{\mathrm{los}}}
\newcommand{\kperp}{k^{\mathrm{perp}}}
\newcommand{\kf}{k_{\mathrm{f}}}

\title[The Bispectrum and 21cm Foregrounds.]
{The bispectrum and 21cm foregrounds during the Epoch of Reionization.}
\author[C. A. Watkinson, C. M. Trott \& Ian Hothi]
{Catherine ~A ~Watkinson$^{1, 2}$\thanks{Email: \href{mailto:catherine.watkinson@gmail.com}
{\protect\nolinkurl{catherine.watkinson@gmail.com}}}, Cathryn ~M ~Trott$^{3, 4}$, Ian Hothi$^1$\\
$^1$Department of Physics, Blackett Laboratory, Imperial College, London, SW7 2AZ, UK \\
$^2$School of Physics and Astronomy, Queen Mary University of London, Mile End Road, London E1 4NS, UK \\
$^3$International Centre for Radio Astronomy Research (ICRAR), Curtin University, Bentley WA, Australia \\
$^4$ARC Centre of Excellence for All Sky Astrophysics in 3 Dimensions (ASTRO 3D), Bentley WA, Australia \\
}

\date{\today}

\pubyear{\number\year}

\begin{document}
\maketitle

\begin{abstract}
Numerous studies have established the theoretical potential of the 21cm bispectrum to
boost our understanding of the Epoch of Reionization (EoR), and therefore early generation of stars and galaxies.
In this paper we take a first look at the impact of foregrounds and instrumental effects on the 21cm bispectrum
and our ability to measure it.
Unlike the power spectrum for which (in the absence of instrumental effects) there is a window clear of smooth-spectrum foregrounds in which
the 21cm power spectrum may be detectable, there is no such "EoR window" for the bispectrum.
Instead, on smaller scales EoR structure modulates that of the foregrounds (FG)
to alter the EoR+FG bispectrum from that of the foregrounds in a complex manner.
On larger scales the EoR structures are completely swamped by those of the foregrounds,
and the EoR+FG bispectrum is entirely dominated by that of the foregrounds.
It is therefore unlikely that the bispectrum will be useful in cases where we are
observationally restricted to using foreground avoidance techniques.
We also find that there is potential for instrumental effects to seriously corrupt the
bispectrum, possibly even rendering the bispectrum
useless for parameter estimation.
On larger scales ($\kperp \le 0.3$ \invmpc), foreground removal using GMCA is found to recover
the EoR bispectrum to a reasonable level of accuracy (even better than the power spectrum for certain configurations).
Further studies are necessary to understand the error and/or bias associated with
foreground removal before the 21cm bispectrum can be practically applied in
analysis of future data.
\end{abstract}

%%%%%%%%%%%%%%%%%%%%%%%%%%%%%%%%%%%%%%%%%%%%%%%%%%%%%%%%%%%%
\section{Introduction}\label{sec:intro}
%%%%%%%%%%%%%%%%%%%%%%%%%%%%%%%%%%%%%%%%%%%%%%%%%%%%%%%%%%%%
Understanding the properties of the first stars and galaxies, as well as their successors,
is of high priority if we are to fully understand our Universe.
Of particular interest is the Epoch of Reionisation (EoR) during which these
early generations of stars and galaxies progressively ionised the otherwise neutral
intergalactic medium and (likely prior to the EoR) heated it during the Epoch of Heating (EoH).
See \citealt{Loeb2013} for a good review of these physical processes.

Many ground breaking radio interferometers have been built
(or are under development) that aim to
map the 21cm line of neutral hydrogen,
either in emission or absorption, as a function of redshift.
Such datasets would ultimately provide three dimensional samples of our Universe
and would revolutionise our understanding.
The 21cm hyperfine transition (produced by a spin flip in the lowest energy of neutral
hydrogen) is sensitive to heating processes as we expect Lyman-alpha coupling to
quickly couple the spin temperature to that of the kinetic gas temperature.
Since the signal only comes from neutral hydrogen, it is also sensitive to the
progress of reionisation.

The first generation of interferometer such as LOFAR\footnote{\sloppy The LOw Frequency ARray \url{http://www.lofar.org/}}, MWA\footnote{The Murchison Wide-field Array
\url{http://www.mwatelescope.org/}} and PAPER\footnote{\sloppy The Precision Array to Probe Epoch of Reionization
\url{http://eor.berkeley.edu/}}
are all hoping to make detections, but are currently limited by unresolved systematics
to placing upper limits on the 21cm power spectrum \citep{Patil2017, Gehlot2018, Li2019a, Kolopanis2019}.
There is hope that they will be able to get on top of systematics in order to
integrate down the noise to make a detection of the EoR power spectrum.
It is also worth noting that experiments to detect the global 21cm signal have also
been developed, most notably EDGES\footnote{\sloppy The Experiment to Detect the Global EoR Signature \url{http://loco.lab.asu.edu/edges/}}
has claimed a detection of an extreme trough in the signal during the Epoch of Heating.
If true this requires us to look beyond our fiducial model for either exotic cooling mechanisms
or for a source of excess background with 21cm wavelength \citep{Bowman2018a}.
Although, it is consensus that another instrument needs to confirm the findings before
we be sure that EDGES has made a genuine detection \citep{Hills2018, Bradley2019, Sims2020}.

It is expected that the next generation of radio interferometer, the SKA\footnote{\sloppy The Square Kilometre Array
\url{http://www.skatelescope.org/}}
and HERA\footnote{\sloppy The Hydrogen Epoch of Reionization Array
\url{http://reionization.org/}},
will lead us into a era of precision high-redshift 21cm observations \citep{Braun2019}.
In preparation for these datasets, a great deal of effort continues to be poured into
simulating the 21cm signal so that we may robustly make sense of observations.
The complexity of the physical processes involved mean there are a great deal of
degeneracy between different parameters involved in simulating the 21cm signal,
see for example and \citealt{Greig2017} and \citealt{Park2018}.

The 21cm signal us predicted to be extremely non-Gaussian throughout the
EoR, it therefore makes sense to consider the gains of using
statistics like the bispectrum that are sensitive to non-Gaussian structure in
data.
The bispectrum is the Fourier dual to the three-point correlation function, which
measures the excess probability of signal as a function of three points in a dataset.
This connection to three physical points in real space enforces the bispectrum to be
a function of three \vect{k} vectors that form a closed triangle.
Many theoretical studies have shown that there is valuable additional information to be gained by
measuring the Bispectrum, e.g \citealt{Shimabukuro2016a, Majumdar2017, Watkinson2018, Hutter2019}.

A major challenge to observing the 21cm line is that there are strong radio foregrounds
at the frequencies of interest that are several orders of magnitude larger than the signal
of interest.
The work of \citealt{Trott2019} also indicates that the bispectrum of certain \vect{k}-triangle configurations
may be less foreground corrupted than the power spectrum, and so exhibit higher
signal to noise.
This result was based on a theoretical model of the bispectrum poisson-distributed
point sources. However, synchrotron and free-free diffuse emissions from our galaxy
and extragalactic diffuse emission account for most of the 21cm foregrounds at the
redshifts of relevance to the EoR ($z\,\sim\, 6-15$) and beyond to the cosmic dawn \citep{Shaver1999b, DiMatteo2004, Gleser2008, Liu2012, Murray2017, Spinelli2018}.

In this work we analyse the bispectrum from foregrounds that exhibit realistic
structure on the sky, the cosmological signal, and the
combination of both.
We also consider how the bispectrum might be impacted by observations using the SKA
by analysing the bispectrum of data that has been passed through the radio
intereferometer simulation \oskar\footnote{\sloppy Observational effects were simulated
using \oskar \url{https://github.com/OxfordSKA/OSKAR} }.
This work is a first step towards understanding how useable the bispectrum will be
in practice for understanding the nature of the first stars and galaxies.

This paper is structured as follows. In Section~\ref{sec:sims} we provide an overview of the simulations
we analyse in this paper and of the algorithm we use to measure the bispectrum.
In Section~\ref{sec:EoRwindow} we consider whether or not there is an equivalent
to the power spectrum EoR window for the bispectrum, we will see that whilst the presence
of EoR structures alter the bispectrum from that of the foregrounds alone, there is not
a clear cut bispectrum EoR window.
In Section~\ref{sec:instrumentals} we take a look at the impact of simulating instrumental
effects on the foreground bispectrum, finding that there is potential for instrumentals
to substantially corrupt the bispectrum from that of the underlying clean signal.
Also in this section, we also consider how much sample variance of the foregrounds might
impact the foreground bispectrum.
We then consider how well foreground removal using GMCA might be able to recover the
clean EoR bispectrum in Section~\ref{sec:GMCA}. We will see that GMCA EoR residuals
exhibit a bispectrum with the correct sign and order of magnitude as that of the clean EoR signal,
but with qualitative and quantitative differences.
Finally, in Section~\ref{sec:Conc} we detail our conclusions.

\section{Overview of simulations and analysis} \label{sec:sims}

\subsection{Foreground simulations}
In this work we utilise the foreground simulations of \citealt{Li2019} (hereafter \textbf{Li2019})
that extrapolate from lower redshift observations to produce
foreground simulations that exhibit realistic structures
on the sky.\footnote{\sloppy The simulations we analyse may be
acquired from \url{https://github.com/liweitianux/cdae-eor} and the
associated foreground simulation package at \url{https://github.com/liweitianux/fg21sim}}
\citealt{Li2019} include all the major contributions to the foregrounds mentioned in the previous section.
Diffuse galactic synchrotron is extrapolated to higher redshifts using the 408 MHz all-sky
Haslam maps as a basis, with substructure simulated by extrapolation of the
galactic-emission power spectrum \citep{Haslam1983, Wang2010}.
Galactic free-free emission is assigned using its close relation to the H$\alpha$ line as
observed by \citealt{Finkbeiner2003} and extrapolating to higher redshifts.
The galactic diffuse emission was simulated at an (R.A., Dec.) = (0$^{\circ}$, -27$^{\circ}$)
which is at a high galactic latitude appropriate for simulating an SKA observation \citep{Beardsley2016a}.
Semi-analytical modeling are employed for the less dominant foreground contributors.
Li2019 assume that point sources with a flux greater than 10~mJy (at 158 MHz) have been successfully removed.
We refer the interested reader to \citealt{Wang2010}, \citealt{Wang2013} and
\citealt{Li2019b} for details of the foreground simulations.

\subsection{Epoch of Reionisation simulations}
The Evolution of 21cm Structure EoR datasets were used to generate the EoR skymaps,
using the faint galaxies model.\footnote{\sloppy The EoR simulations used were the faint galaxies model from \url{http://homepage.sns.it/mesinger/EOS.html} } These were tiled and resampled to match the
foreground simulation's resolution, namely ($1800^2$) pixels with a field of view 10$^{\circ}$ by 10$^{\circ}$, giving a resolution of 20 arcsecs.
The Li2019 datasets consist of 101 frequency slices over the frequency range
154--162 MHz. This frequency range corresponds to $z = 7.77 - 8.22$ and a channel width of 80 kHz;
the central frequency is 158 MHz observing at $z = 8$ at which the neutral fraction
of the EoR simulation is roughly 50\%.

\subsection{Simulating observational effects}
\begin{figure*}
\begin{minipage}{176mm}
\begin{tabular}{c}
  \includegraphics[trim=1.5cm 0cm 0.5cm 0.0cm, clip=true, scale=0.46]{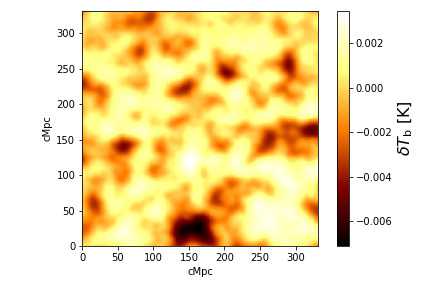} \includegraphics[trim=2cm 0cm 0.5cm 0.0cm, clip=true, scale=0.46]{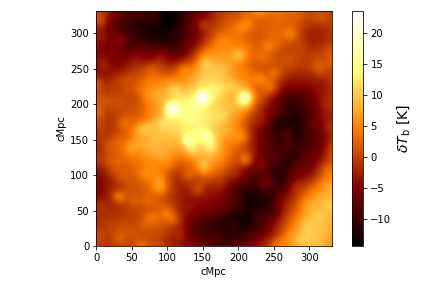} \includegraphics[trim=2cm 0cm 0.5cm 0.0cm, clip=true, scale=0.46]{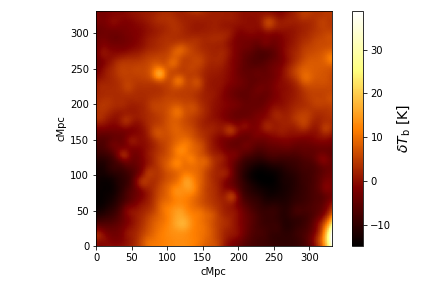}\\
\end{tabular}
\caption{154 MHz slice of the Li2019 datasets including simulated observations by SKA-LOW phase 1.
The left plot shows the "observed" EoR signal, the centre plot shows the "observed" foregrounds in field 1,
and the right plot shows the "observed" foregrounds in field 2.
These all correspond to a field-of-view of 2$^\circ$.
}\label{fig:obs_slices}
\end{minipage}
\end{figure*}

The foreground and EoR skymaps were all passed through the \oskar simulation and
then imaged using \wsclean\footnote{\sloppy Imaging was done using \wsclean
\url{https://sourceforge.net/p/wsclean} }
with natural weighting \citep{Offringa2014}
assuming the current design for the SKA-Low phase 1 telescope model;
i.e. 224 stations (each of diameter 35m containing 256 randomly placed antennas)
randomly distributed in a core of diameter 1~km, with the remains
occupying three spiral arms that extend out to a radius of 35~km.\footnote{\sloppy SKA1-Low
layout used for the \oskar telescope model was based on
\url{https://astronomers.skatelescope.org/wp-content/uploads/2016/09/SKA-TEL-SKO-0000422_02_SKA1_LowConfigurationCoordinates-1.pdf} }
They simulate a 6-hour tracked observation to acheive a full $uv$-sampling and
exclude noise to be consistent with a 1000 hour SKA integration time for $z \sim 8$.
The final datasets analysed here are cropped from the simulated observation to have
a 2$^{\circ}$ by 2$^{\circ}$ FoV of $360^2$ pixels.
Li2019 also follow the above procedure for an (R.A., Dec.) = (3$^{\circ}$, -27$^{\circ}$).
We will refer to this simulated observation as "observed" \textbf{field 2} and the R.A. = 0$^{\circ}$
simulated observations as "observed" \textbf{field 1}.
Slices from the "observed" field 1, field 2, and the EoR signal are shown in Fig.~\ref{fig:obs_slices}
for the 154~MHz slice.

\begin{figure}%[!hbtp]
\centering
  $\renewcommand{\arraystretch}{-0.75}
  \begin{array}{c}
    \includegraphics[trim=0.cm 0.0cm 20.5cm 1.5cm, clip=true, scale=0.4]{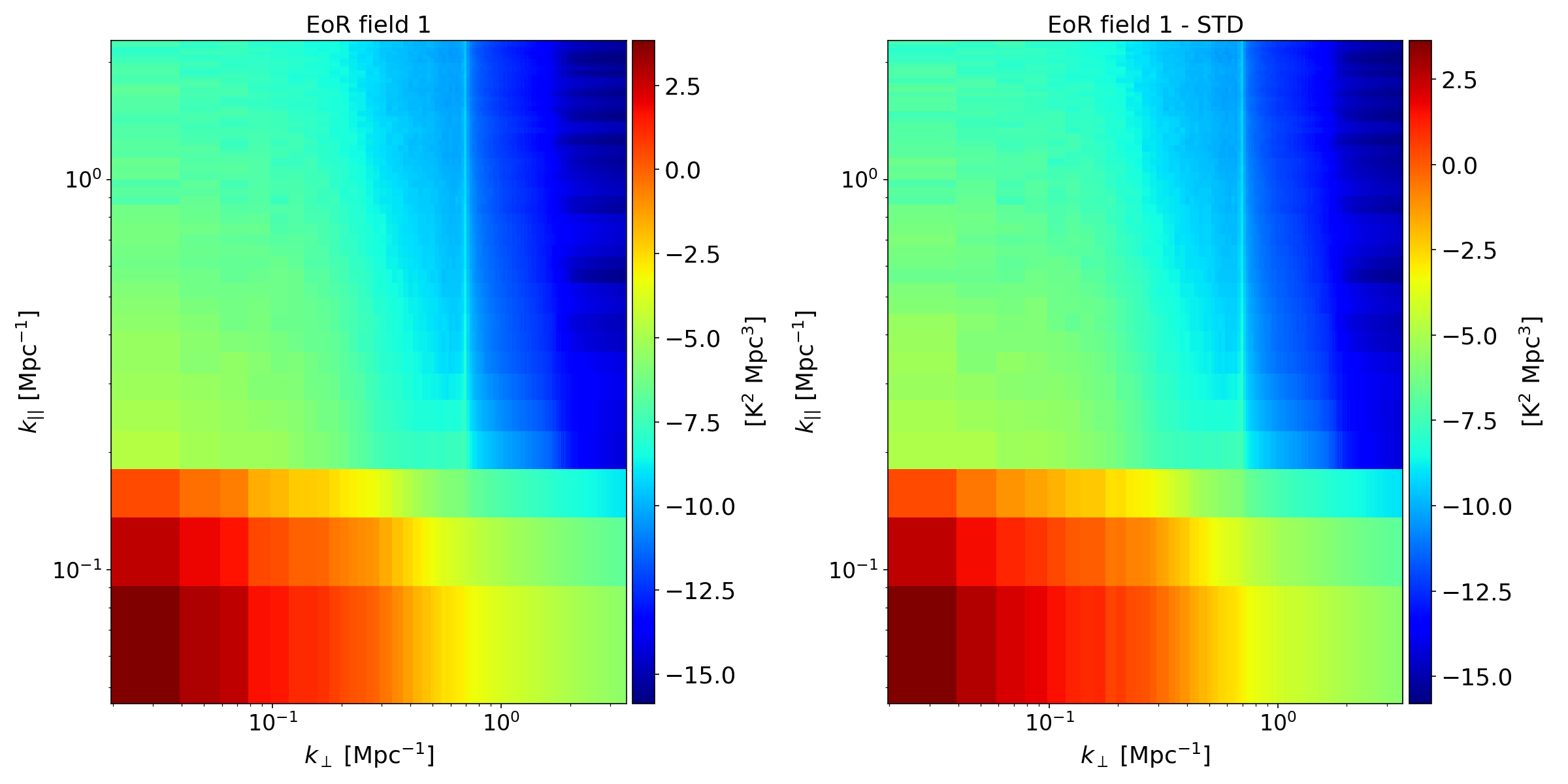}\\
  \end{array}$
  \caption{Power spectrum from the "observed" EoR+FG (field 1) dataset.
  for $\klos>0.2$ \invmpc the power spectrum is dominated by both the galactic emission and EoR
  signal. We see an artifact at $\kperp\sim 0.7$ \invmpc and suppression
  of power for $\kperp> 0.7$ \invmpc. As such, we conservatively restrict our analysis to $k$ vectors
  with $k < 0.6$ \invmpc}
  \label{fig:2Dpspec}
\end{figure}

Following works like Li2019 and \citealt{Chapman2016}, we apply an extended Blackman-Nuttall
filter over the frequency axis to suppress non-periodicity issues when Fourier Transforming the data.
We will also restrict our analysis to $k < 0.6$ \invmpc because
there is an artifact at this scale and a suppression of power beyond.
This can be seen in Fig.~\ref{fig:2Dpspec}, where we plot the power spectrum
as a function of scale on the sky ($\kperp$ or $k_\perp$) and in the line of sight ($\klos$ or $k_\parallel$).
This artifact is produced at the imaging stage because Li2019 make a $uv$ cut at 1000 $\lambda$,
effectively erasing the power on smaller scales and therefore achieving a resolution similar to smaller arrays such as the MWA.
This can be alleviated by using a Briggs weighting with a higher cutoff in baseline length.
However, it is sufficient for studying EoR structures which are on larger scales.

%A similar artifact is also seen in \citealt{Chapman2016}. \citealt{Chapman2016} attributes this to the LOFAR point-spread function, and Li2019 to their having excised the lowest 6 Fourier coefficients to suppress the foreground emission. However, it is our intuition, that neither effect should produce such a sharp feature in the power spectrum. We therefore suggest that it is possible that this feature is in fact an edge effect, where the wings of the primary beam (or side lobes) are observing the hard edge of the square simulated cosmological sky model. Looking at the simulated SKA beam shapes in Figure 15 of \citealt{DeLeraAcedo2018}, this seems quite plausible. If we are correct, then this effect can be alleviated by instead using Briggs weighting when imaging, as done by \citealt{Li2019b}, who do not see such an artifact, despite using an identical telescope model, simulation resolution and 10$^\circ$ FoV. We will investigate this point further in a follow up paper.

\subsection{Measuring the cylindrically-averaged bispectrum}
The bispectrum measures the level of correlation between three different scales
defined by three $k$ vectors $\boldsymbol{k}_1, \boldsymbol{k}_2, \boldsymbol{k}_3$
that form a closed triangle. The bispectrum is the Fourier dual to the real-space three-point correlation
function (the excess probability as a function of three points) and is defined as,
\begin{equation}
\begin{split}
(2\pi)^3 B(\vect{k}_1, \vect{k}_2, \vect{k}_3) \delta^{\mathrm{D}}(\vect{k}_1 + \vect{k}_2 +  \vect{k}_3  )=
\langle \Delta(\vect{k}_1)\Delta(\vect{k}_2)\Delta(\vect{k}_3)\rangle \,,\\
\label{eq:bi_definition}
\end{split}
\end{equation}
Where $\delta^{\mathrm{D}}(\vect{k}_1 + \vect{k}_2 + \vect{k}_3)$ is the Dirac-delta function.

The bispectrum provides some sensitivity to the presence of structure in a map,
unlike the power spectrum which is unable to distinguish a dataset with structure
from a Gaussian random field.
The interference pattern of the three plane waves associated with $\vect{k}_1, \vect{k}_2, \vect{k}_3$
for a given triangle configuration show the types of structure the bispectrum
for that configuration is sensitive to.
For example, the interference pattern of an equilateral triangle
consists of regularly spaced filaments of above average signal with a circular cross section.
The more the signal in a dataset follows this interference pattern, the stronger the equilateral
bispectrum will be relative to other configurations.
Squashing the equilateral triangle configuration, so that one leg of the triangle is longer than
the other two, squashes the circular cross sections into elliptical cross sections; at the extreme
becoming almost planar.
Squeezing the equilateral configuration so that one leg of the triangle is smaller
modulates a large scale mode over a smaller scale interference pattern.
Should the structure in a map be driven by concentrations of below-average signal,
rather than above-average signal, the bispectrum will be negative.
We shall consider such interference patterns in more depth in Section~\ref{sec:EoRwindow},
but also see \citealt{Lewis2011, Watkinson2018, Hutter2019} for more discussion of
how to interpret of the bispectrum.

The bispectrum results in this paper are performed using an adapted version of
the code described in \citealt{Watkinson2017} which exploits Fast-Fourier Transforms (FFTs)
to enforce the Dirac-delta function to efficiently measure the spherically-averaged bispectrum.
The baselines of a radio interferometer sample $uv$-space (which is linearly related to $\vect{k}$ space).
An observation by such an instrument will therefore produce a sampling of $\vect{k}^{\mathrm{perp}}$,
i.e. the $\vect{k}$ modes across the sky, for each frequency channel it observes.
A Fourier Transform can then be performed in the frequency axis to produce a sampling
of $\vect{k}^{\mathrm{los}}$, i.e. the $\vect{k}$ modes along the line of sight.
See \citealt{Morales2004} for the equations that connect the telescope observing co-ordinates
to $\vect{k}$ in inverse comoving Mpc (\invmpc).
It is common to beat down statistical and instrumental noise by studying the spherically-averaged bispectrum, exploiting the fact that we expect the Universe to be homogeneous and isotropic.
However, because we expect the foreground power to be confined to large line-of-sight scales, it is essential to study the observed bispectrum as a cylindrically-averaged quantity.

The code of \citealt{Watkinson2017} estimates $B(k_1, k_2, k_3)$
(the spherically-averaged bispectrum) in the following manner:
\begin{packed_enum}
\item FFT the dataset to $d^{ \mathrm{fft} }$;
\item from this create three new masked datasets: $d_1^{ \mathrm{fft} }$ containing the $d^{ \mathrm{fft} }$
values in a spherical shell whose $|\vect{k}| \sim k_1$ and zero otherwise,  $d_2^{ \mathrm{fft} }$ containing the $d^{ \mathrm{fft} }$
values in a spherical shell whose $|\vect{k}| \sim k_2$ and zero otherwise, and  $d_3^{ \mathrm{fft} }$ containing the $d^{ \mathrm{fft} }$
values in a spherical shell whose $|\vect{k}| \sim k_3$ and zero otherwise;
\item create three other masked datasets $I_1^{ \mathrm{fft} }$, $I_2^{ \mathrm{fft} }$, $I_3^{ \mathrm{fft} }$
as in the previous step, but with 1's instead of the $d^{ \mathrm{fft} }$ values;
\item perform an inverse FFT on the $d_i^{ \mathrm{fft} }$ and $I_i^{ \mathrm{fft} }$ to produce $d_i'$ and $I_i'$
\item estimate $B(k_1, k_2, k_3)$ by summing over all pixels $(d_1'\,d_2'\,d_3')/ (I_1'\,I_2'\,I_3')$
(applying a piecewise product) and applying Fourier normalisations.
\end{packed_enum}

There are many ways in which steps (ii) and (iii) of this process might be adapted to instead measure the
cylindrically-averaged bispectrum, all relating to how you choose to bin the triangle vertices.
After various tests we find that the optimal way is to use a binwidth of a fixed
number of pixels when deciding whether a given pixel meets the requirement $|\vect{k}| \sim k_i$ for $i=1,2$;
i.e. $|\vect{k}| = k_i \pm \Delta k$ where $\Delta k = n \kf$, $n$ is an integer,
$\kf = 2\pi/L$ is the fundamental pixel size in $\vect{k}$ space, and $L$ is the physical size
of the simulation side.

Then in addition we bin the closure vector's magnitude $|\vect{k}_3|$ by $\pm 0.05\,\theta_{12}$
where $\theta_{12}$ is the angle between $\vect{k}_1$ and $\vect{k}_2$.\footnote{Our choice for the binning
of the closure vector $|\vect{k}_3|$ is motivated by previous tests in \citealt{Watkinson2017} and \citealt{Watkinson2018}}
Finally, we cut the spherical shells of the $d_i^{ \mathrm{fft} }$ and $I_i^{ \mathrm{fft} }$
into rings according to the $\klos$ for each $\vect{k}_i \pm \Delta k'$, where we include the prime
to allow for the binning in the line-of-sight to be different from that used for calculating
$|\vect{k}|$.

In tests using \cmfast simulations of the EoR signal (see \citealt{Signal2010}
for details of \cmfast) with different resolutions and initial conditions,
we found that using $\pm 4$ pixels, i.e. $\Delta k = 4 \kf$, provides optimal stability to sample variance.
However, because the line-of-sight fundamental scale $k_{\mathrm{f}}$ is relatively large due to the bandwidth of the Li2019 dataset
($k_{\mathrm{f}} = 0.05$ \invmpc), we bin $\klos$ by $\pm 2 \kf$ and $\kperp$
by $\pm 4 \kf$, i.e $\Delta k = 4 \kf$ and $\Delta k' = 2 \kf$ .
This is reasonable in this case as we only consider one realisation for the EoR.
However, it would be better to ensure a finer line-of-sight $k$ binning that allows for
$\pm 4 \kf$ in any analysis that attempts to perform parameter estimation or similar.
However, the evolution of the signal will become an issue for bandwidths much larger than 10 MHz.

\begin{figure}%[!hbtp]
\centering
  $\renewcommand{\arraystretch}{-0.75}
  \begin{array}{c}
    \includegraphics[trim=0.cm 9.65cm 0.0cm 0.0cm, clip=true, scale=0.6]{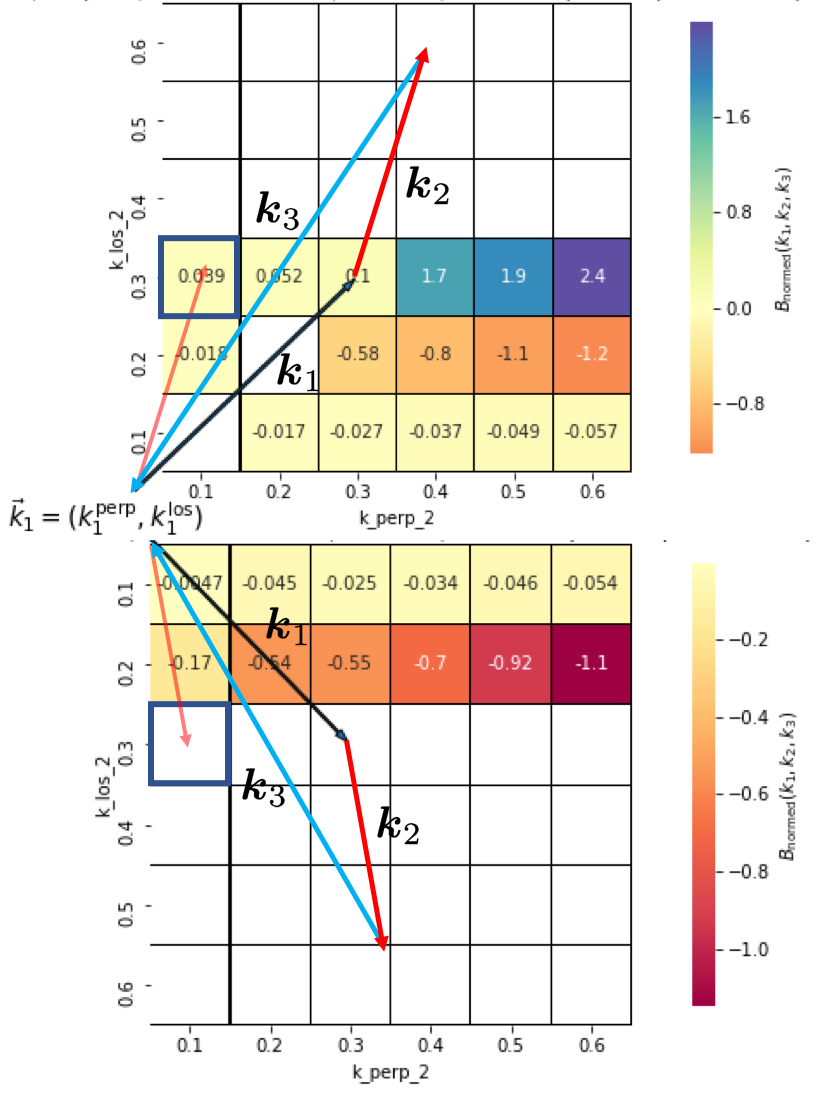}\\
  \end{array}$
  \caption{Schematic of how to interpret the triangle configurations for a particular
  square in $(\kperp_2, \klos_2)$ space.
  The black squares indicate the bispectrum square that the illustrated $\boldsymbol{k}_1$, $\boldsymbol{k}_2$,
  $\boldsymbol{k}_3$ vector triangle corresponds to.}
  \label{fig:tri_interp}
\end{figure}

We present our bispectrum results as a function of $\kperp_2$ and $\klos_2$
with each plot corresponding to a particular $k_1 = (\kperp_1, \klos_1)$.
We annotate an arrow onto all plots to provide a visual description of $\vect{k}_1$
and each coloured square in our plots corresponds to a distinct triangle configuration.
Fig.~\ref{fig:tri_interp} illustrates how to interpret which $\boldsymbol{k}_1$, $\boldsymbol{k}_2$,
$\boldsymbol{k}_3$ vector triangle configuration
a given square (in the figure denoted by a black box) corresponds to.
The vector for $\boldsymbol{k}_2$ is found by drawing an arrow from the origin of the $\boldsymbol{k}_1$ vector to the square of interest.
This arrow is the $\boldsymbol{k}_2$ vector associated with the square.
The triangle probed by this square can then be formed by transforming this arrow
so that its origin follows from the end of the $\boldsymbol{k}_1$ arrow.
For every plot of $k_1 = (\kperp_1, \klos_1)$ we show,
we have examined three other plots corresponding to
to the set of triangles associated with $(\kperp_1, -\klos_1)$, $(-\kperp_1, \klos_1)$,
and $(-\kperp_1, -\klos_1)$.
As in Fig.~\ref{fig:tri_interp},
we will only show the $k_1 = (\kperp_1, \klos_1)$ plots throughout this paper as they
sufficient to illustrate the key points that are raised in our analysis.
We have also considered a range of $k_1 = (\kperp_1, \klos_1)$, but focus the presented
analysis on $(\kperp_1 = 0.1$ \invmpc, $\klos_1 = 0.1$ \invmpc) and $(\kperp_1 = 0.3$ \invmpc, $\klos_1 = 0.3$ \invmpc)
as again, these illustrate the main points we wish to convey in this paper
as it is the foreground $\klos_1$ that dominates the effects we are mainly
interested in here.\footnote{In our full analysis we also considered $(\kperp_1 = 0.1$ \invmpc, $\klos_1 = 0.3$ \invmpc)
and $(\kperp_1 = 0.3$ \invmpc, $\klos_1 = 0.1$ \invmpc)}

Throughout this work we present analysis using a normalised version of the bispectrum
that is common in signal processing as it isolates the non-Gaussianity of the phases:

\begin{equation}
b(k_1, k_2, k_3) = \frac{B(k_1, k_2, k_3)}{\sqrt{(k_1\,k_2\,k_3)^{-1}\,P(k_1)\,P(k_2)\,P(k_3)}}\,.\\
\label{eqn:normB}
\end{equation}
\citealt{Watkinson2018} find this to be the best normalisation for interpreting
the 21cm bispectrum.
For the rest of this paper, when we refer to the bispectrum, we are referring to the
normalised bispectrum of Eqn. \ref{eqn:normB}.
It is worth noting that there are potentially issues in using this statistic
in practice if there are differences in the way that foreground residuals (or instrumental effects) propagate
onto the power spectrum and bispectrum estimators \citep{Trott2019}.

%%%%%%%%%%%%%%%%%%%%%%%%%%%%%%%%%%%%%%%%%%%%%%%%%%%%%%%%%%%%%%%%%%%%%%%%%%%%%%%%
%%%%%%%%%%%%%%%%%%%%%%%%%%%%%%%%%%%%%%%%%%%%%%%%%%%%%%%%%%%%%%%%%%%%%%%%%%%%%%%%
\section{Is there an EoR window for the bispectrum?} \label{sec:EoRwindow}
%%%%%%%%%%%%%%%%%%%%%%%%%%%%%%%%%%%%%%%%%%%%%%%%%%%%%%%%%%%%%%%%%%%%%%%%%%%%%%%%
%%%%%%%%%%%%%%%%%%%%%%%%%%%%%%%%%%%%%%%%%%%%%%%%%%%%%%%%%%%%%%%%%%%%%%%%%%%%%%%%
Foregrounds at the frequencies relevant to 21cm observations are all expected to
get gradually stronger with decreasing observational frequency.
With power coming from such a large-scale frequency modes it is expected that foreground power will
be restricted to small $\klos$, with the chromatic nature of the instrument smearing
some of the foreground power into a wedge like feature \citep{Datta2010, Vedantham2012, Morales2012, Thyagarajan2013, Hazelton2013, Liu2014c}.
As such, there should be an EoR window largely clean of foregrounds
in which the power spectrum from the EoR will dominate.
Although, \citealt{Li2019b} show that the presence of radio haloes could drastically reduce
the signal to noise for $k<1$ \invmpc even within the EoR window.
In this section we will consider whether or not such a window exists for the bispectrum;
we analyse the bispectrum from the EoR signal (\textbf{EoR-only}),
the foreground signal (\textbf{FG-only}) and the combined field (\textbf{EoR+FG}).
We emphasise that the bispectrum of the EoR+FG signal cannot be considered as a simple sum of the EoR-only and FG-only bispectra.
This is because the EoR structures modulate with the foreground structures.
On top of this, the bispectra we present here are normalised by the power spectrum to isolate the non-Gaussianity in the phases.
This further complicates a simple propagation as the denominator and numerator both change between the fields.

\begin{figure}%[!hbtp]
\centering
  $\renewcommand{\arraystretch}{-0.75}
  \begin{array}{c}
    \includegraphics[trim=10.9cm 0.0cm 4.0cm 1.5cm, clip=true, scale=0.5]{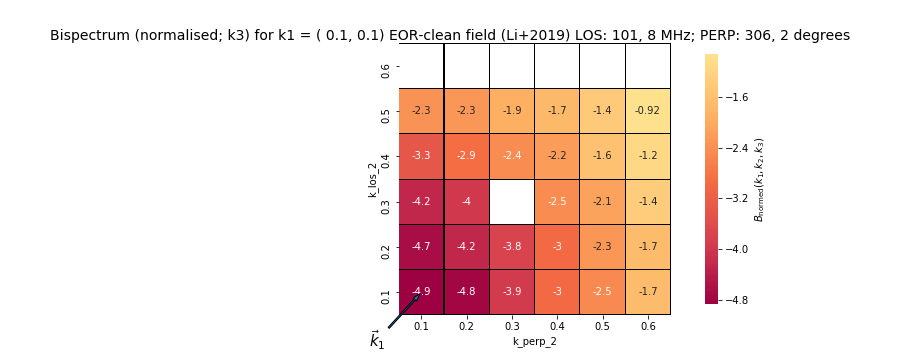} \\
    \includegraphics[trim=10.9cm 0.0cm 4.0cm 1.5cm, clip=true, scale=0.5]{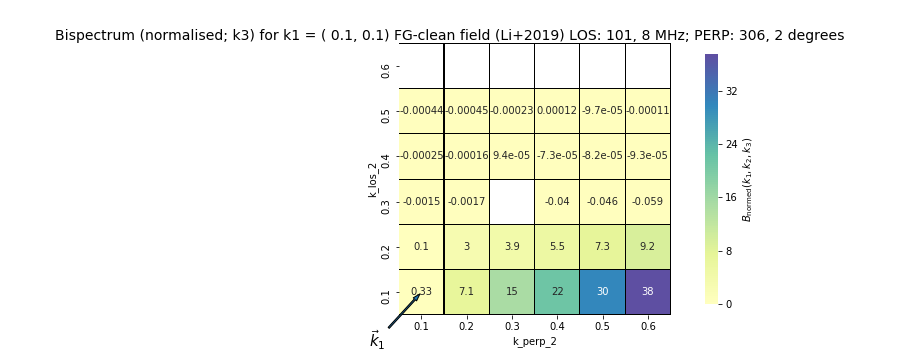} \\
    \includegraphics[trim=10.9cm 0.0cm 4.0cm 1.5cm, clip=true, scale=0.5]{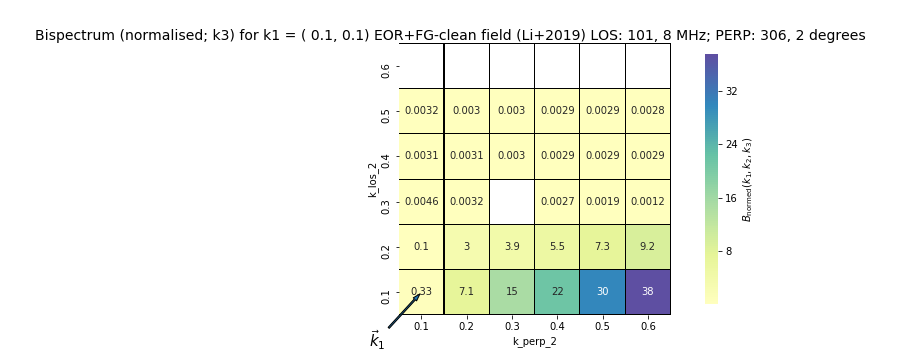}\\
  \end{array}$
  \caption{Normalised bispectrum from clean
  field 1 for $(\kperp_1, \klos_1) = (0.1$ \invmpc, $0.1$ \invmpc) and for the EoR signal (EoR-only; top),
  the foregrounds signal (FG-only; middle) and for the combined field (EoR+FG; bottom).
  We plot this as a function of $\kperp_2$ and
  $\klos_2$ so that each square corresponds to a different $\vect{k}$-triangle configuration.
  There is no evidence of a clean window in which the EoR bispectrum dominates.}
  \label{fig:kperp_pt1_clean}
\end{figure}

Before we consider instrumental effects on the bispectrum we study the clean simulations
so we may detangle which features are inherent and which relate to instrumental effects.
In Fig.~\ref{fig:kperp_pt1_clean} we present the cylindrically-averaged bispectrum
of the EoR-only (top),
FG-only (middle) and EoR+FG (bottom) all from clean field 1, for $(\kperp_1, \klos_1) = (0.1$ \invmpc, 0.1 \invmpc).
On these larger-scales, the foreground bispectrum is always positive, as predicted
by \citet{Trott2019} for smooth-spectrum point sources.
Despite the FG-only bispectrum being well confined to $\klos_2\le 0.2$ \invmpc and the EoR-only
bispectrum being non-negligible and negative at $\klos_2> 0.2$ \invmpc, the EoR structure has no discernible
influence on the EoR+FG bispectrum for all triangle configurations for $(\kperp_1, \klos_1) = (0.1$ \invmpc, 0.1 \invmpc).

\begin{figure}%[!hbtp]
\centering
  $\renewcommand{\arraystretch}{-0.75}
  \begin{array}{c}
    \includegraphics[trim=10.9cm 0.0cm 4.0cm 1.5cm, clip=true, scale=0.5]{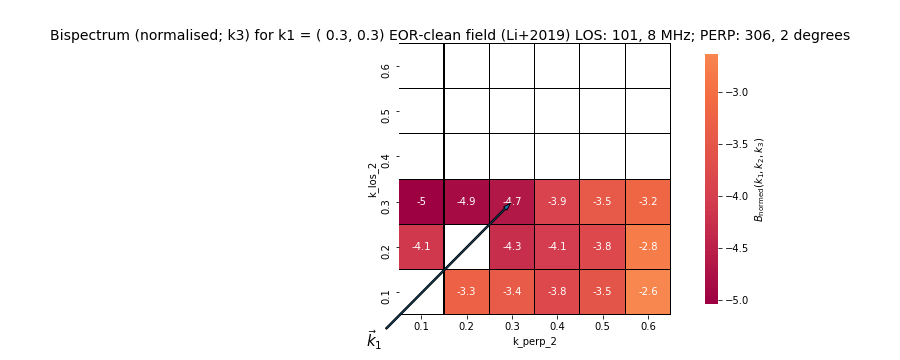} \\
    \includegraphics[trim=10.9cm 0.0cm 4.0cm 1.5cm, clip=true, scale=0.5]{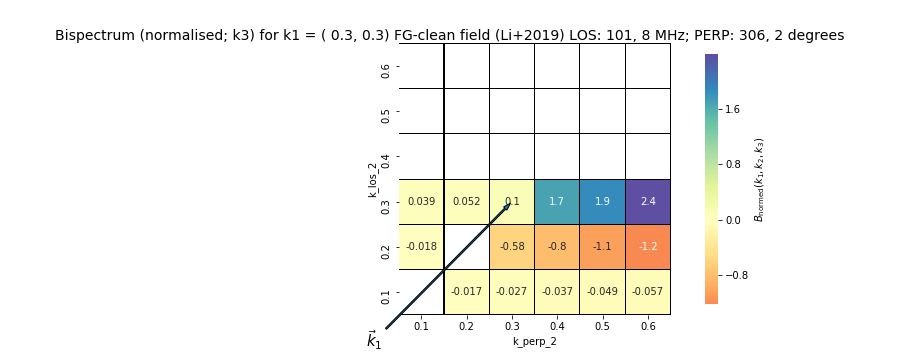} \\
    \includegraphics[trim=10.9cm 0.0cm 4.0cm 1.5cm, clip=true, scale=0.5]{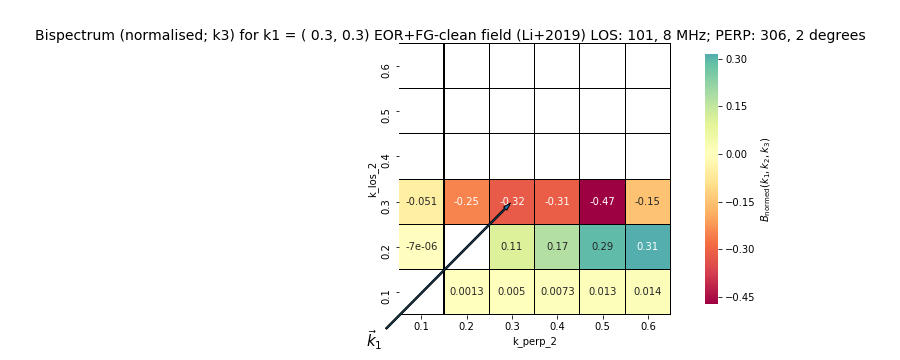}\\
  \end{array}$
  \caption{Normalised bispectrum from clean
  field 1 for $(\kperp_1, \klos_1) = (0.3$ \invmpc, 0.3 \invmpc) and for the EoR signal (EoR-only; top),
  the foregrounds signal (FG-only; middle) and for the combined field (EoR+FG; bottom).
  We plot this as a function of $\kperp_2$ and
  $\klos_2$ so that each square corresponds to a different $\vect{k}$-triangle configuration.
  Whilst the presence of EoR structures does alter the bispectrum from that of the FG-only field,
  it is difficult to interpret as
  there is no clean window in which the EoR bispectrum dominates.}
  \label{fig:kperp_pt3_clean}
\end{figure}

In Fig.~\ref{fig:kperp_pt3_clean} we present the cylindrically-averaged bispectrum
of the EoR-only (top),
the FG-only (middle) and the EoR+FG (bottom) for $(\kperp_1, \klos_1) = (0.3$ \invmpc, 0.3 \invmpc).
On these smaller scales the story is slightly different;
voids in the FG signal mean that on certain scales (e.g. $\klos_2 = 0.2$ \invmpc in Fig.~\ref{fig:kperp_pt3_clean}) the bispectrum
can be negative, which goes against the prediction of \citet{Trott2019} (which only considered smooth-spectrum point source foregrounds).
The bispectrum of EoR+FG (bottom) is also clearly altered from that of FG-only (middle) for these smaller-scale configurations,
but equally exhibits no clear correlation with EoR-only bispectrum (top).
For many configurations (e.g. $\klos_2 = 0.3$ \invmpc in Fig.~\ref{fig:kperp_pt3_clean})
the sign is reversed from positive to negative,
presumably driven by destructive interference
of the below-average signal from ionised regions in the EoR signal with the
above-average foreground structures.
Interestingly, for configurations with negative FG-only bispectrum,
the sign of the EoR+FG bispectrum can positive,
even when the EoR-only bispectrum is also negative.
The foreground bispectrum is only weakly negative on these scales, so presumably
there is also above-average signal present for these configurations in both the FG-only
and the EoR-only datasets that constructively interferes and comes to dominate the
EoR+FG bispectrum.
On these smaller scales, the bispectrum of the foregrounds is smaller than it is for
$(\kperp_1, \klos_1) = (0.1$ \invmpc, 0.1 \invmpc).
We conclude that the structure of the EoR gets washed out in the presence of
strong foreground non-Gaussianities,
but when the foreground bispectrum is comparable in magnitude to that of the EoR signal,
a modulation effect can be seen.
So whilst there is no clear cut EoR window for the bispectrum,
there may still be a way to use the bispectrum for detection verification even if
foreground removal is not possible.
This of course assumes that we are confident that we understand the foreground bispectrum well.
However, as we will see in the next section, instrumental effects may make even this
intractable.

\begin{figure*}
\begin{minipage}{176mm}
\begin{tabular}{c}
  \includegraphics[trim=2.65cm 0.5cm 3.5cm 1.0cm, clip=true, scale=0.47]{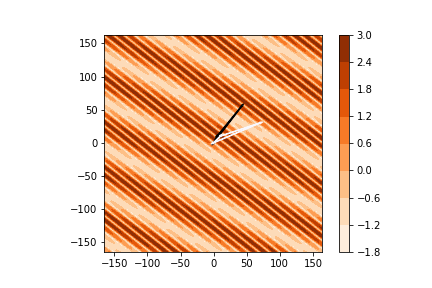} \includegraphics[trim=3.5cm 0.5cm 3.5cm 1.0cm, clip=true, scale=0.47]{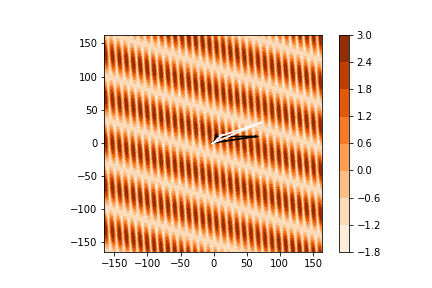} \includegraphics[trim=3.5cm 0.5cm 3.5cm 1.0cm, clip=true, scale=0.47]{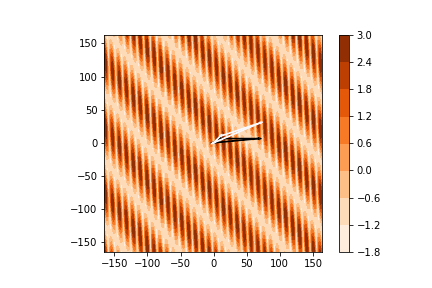} \includegraphics[trim=3.5cm 0.5cm 1cm 1.0cm, clip=true, scale=0.47]{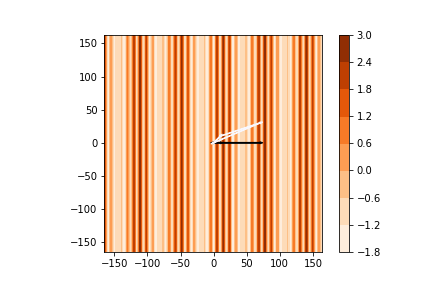}\\
  \includegraphics[trim=1.5cm 0.5cm 3.6cm 1.0cm, clip=true, scale=0.39]{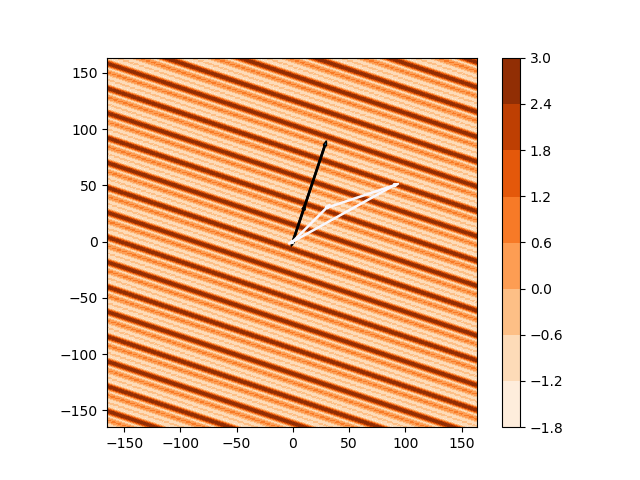} \includegraphics[trim=2.66cm 0.5cm 3.6cm 1.0cm, clip=true, scale=0.39]{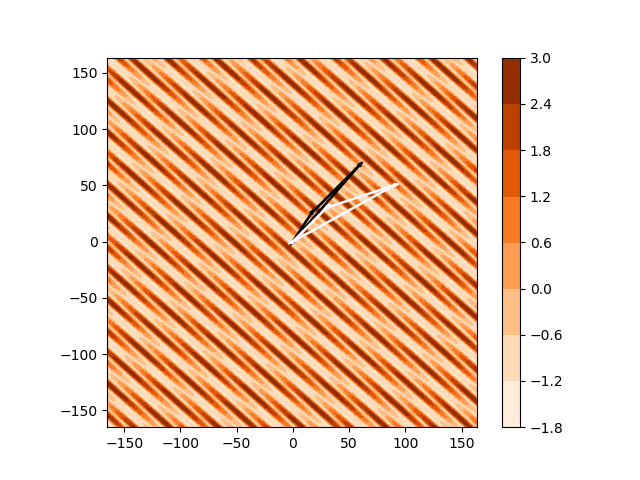} \includegraphics[trim=2.65cm 0.5cm 3.6cm 1.0cm, clip=true, scale=0.39]{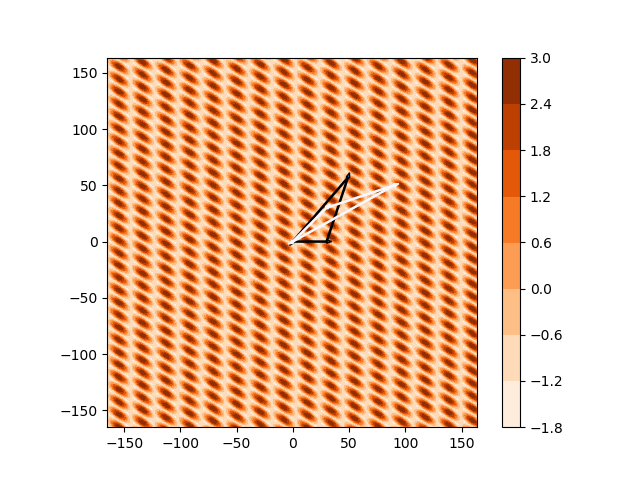} \includegraphics[trim=2.65cm 0.5cm 1cm 1.0cm, clip=true, scale=0.39]{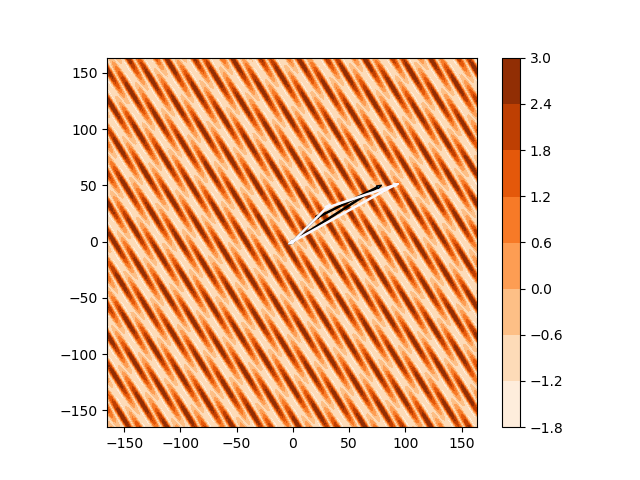}\\
\end{tabular}
\caption{Interference patterns (projected on the perpendicular sky plane) for the waves associated with the Fourier transform
for three $k$ vectors that form a triangle.
We show two configurations that exhibit an extreme foreground bispectrum, one on each row.
Each column corresponds to different $k_x$ and $k_y$ combinations that form
$\vect{k}^{\mathrm{perp}}_1$ and $\vect{k}^{\mathrm{perp}}_2$.
The white triangle shows the shape of the configuration and the
black triangle shows its projection onto the perpendicular sky plane (both scaled up by 100).
The top row corresponds to the $(\kperp_1, \klos_1) = (0.1$ \invmpc, 0.1 \invmpc) and
$(\kperp_2, \klos_2) = (0.6$ \invmpc, 0.1 \invmpc) configuration
and the bottom row to the $(\kperp_1, \klos_1) = (0.3$ \invmpc, 0.3 \invmpc) and $(\kperp_2, \klos_2) = (0.6$ \invmpc, 0.2 \invmpc) one.
}
\label{fig:interference}
\end{minipage}
\end{figure*}

Whilst this paper is less concerned with gaining an in depth understanding of what
structures are driving the bispectrum than we have been in previous works, such as
\citet{Watkinson2018}, it is still interesting to consider what the interference
patterns look like for configurations that exhibit a strong FG bispectrum.
The Fourier transform for a particular $\vect{k}$
is associated with a plane wave through e$^{i \vect{k}x}$;
as such, we can generate an interference pattern of the real parts of the three FFT waves
of a given triangle configuration of $k$-vectors.
This inteference pattern informs us as to what structures a particular $\vect{k}$ triangle probes.
The closer the structure in a dataset to a $\vect{k}$-triangle's interference pattern, the stronger the bispectrum will be for that configuration.
It is important to note that a given structure in a dataset will produce a non-zero bispectrum for many different combinations of $\vect{k}_1, \vect{k}_2, \vect{k}_3$; i.e. there is not a one-to-one relation between the bispectrum for a particular $\vect{k}$-configuration and the real-space structures of your data.

Previous works have presented and discussed the real-part of the interference patterns
for particular $\vect{k}$-triangle shapes, which is useful for interpretting
the spherically-averaged bispectrum from a homogeneous and isotropic dataset \citep{Lewis2011, Watkinson2018, Hutter2019}.
However, given that we are working with our $\vect{k}$ vectors parametrised into
their perpendicular and line-of-sight components, the interpretation of this is
more complex. In particular, for the foregrounds, the structure on the sky does
not change as we scan through in the line-of-sight (although the intensity of the foregrounds
does increase with decreasing frequency).
We have therefore created an animation that scans through the various combinations for the $x$ and $y$
components of $\kperp_1$ and $\kperp_2$, for any given $(\kperp_1, \klos_1)$
and $(\kperp_2, \klos_2)$, and projects the
associated interference pattern onto the perpendicular $x$-$y$ plane.

In Fig.~\ref{fig:interference}, we show a range of interference patterns from this animation for two different configurations.
We include a range of $k_x$ and $k_y$ combinations to illustrate the different types of sky-structure probed by this configuration.
The white triangles show the true shape of this triangle configuration,
and the black triangle its projection onto the perpendicular $x$-$y$ plane;
both are scaled up by a factor of 100 from their natural scales.\footnote{The amplitude ($A$)
of each of the 3D plane waves that we sum to produce the interference patterns
in Fig.~\ref{fig:interference}, is described by $A = \cos( k_x x +  k_y y +  k_y z)$.}
Darker red shading denotes concentrations of above average signal, or positive amplitude.
In the top row of Fig.~\ref{fig:interference} we show a range of interference patterns
for the $(\kperp_1, \klos_1) = (0.1$ \invmpc, 0.1 \invmpc) and $(\kperp_2, \klos_2) = (0.6$ \invmpc, 0.1 \invmpc) configuration
(for which the FG-only bispectrum is strong and positive, as seen in the middle plot of Fig.~\ref{fig:kperp_pt1_clean}).
We see that there is a modulation of a large scale mode over a small scale mode;
this produces bands of tightly-packed ellipses of above-average signal.
Looking at the image of the FG-only data (middle image of Fig.~\ref{fig:obs_slices}),
we see bands of strong FG emission containing compact FG sources surrounded by voids of low FG emission.
It is these features that drive the strong positive bispectrum for such configurations.
In the bottom plot of Fig.~\ref{fig:interference}, we show a range of interference patterns
for the $(\kperp_1, \klos_1) = (0.3$ \invmpc, 0.3 \invmpc) and $(\kperp_2, \klos_2) = (0.6$ \invmpc, 0.2 \invmpc) configuration,
for which the FG-only bispectrum is weekly negative (see the middle plot of Fig.~\ref{fig:kperp_pt3_clean})
and therefore is driven by voids of FG emission.
As seen in the bottom plot of Fig.~\ref{fig:kperp_pt3_clean}, the presence of EoR structure
modulates the FG structure to produce a slightly positive bispectrum for this configuration.
This configuration features regularly spaced filaments of above average signal with mildly elliptical cross sections.
Whilst it is harder to connect this with features in the slices we show in Fig.~\ref{fig:obs_slices},
it is clear that there are many smaller concentrations of above-average signal in the EoR-only
dataset (left image of Fig.~\ref{fig:obs_slices}) that could dominate the bispectrum
for this configuration when combined with the FG-only dataset.

%%%%%%%%%%%%%%%%%%%%%%%%%%%%%%%%%%%%%%%%%%%%%%%%%%%%%%%%%%%%%%%%%%%%%%%%%%%%%%%%
%%%%%%%%%%%%%%%%%%%%%%%%%%%%%%%%%%%%%%%%%%%%%%%%%%%%%%%%%%%%%%%%%%%%%%%%%%%%%%%%
\section{Instrumental effects on the 21cm bispectrum} \label{sec:instrumentals}
%%%%%%%%%%%%%%%%%%%%%%%%%%%%%%%%%%%%%%%%%%%%%%%%%%%%%%%%%%%%%%%%%%%%%%%%%%%%%%%%
%%%%%%%%%%%%%%%%%%%%%%%%%%%%%%%%%%%%%%%%%%%%%%%%%%%%%%%%%%%%%%%%%%%%%%%%%%%%%%%%

\begin{figure}%[!hbtp]
\centering
  $\renewcommand{\arraystretch}{-0.75}
  \begin{array}{c}
    \includegraphics[trim=10.9cm 0.0cm 4.0cm 1.5cm, clip=true, scale=0.5]{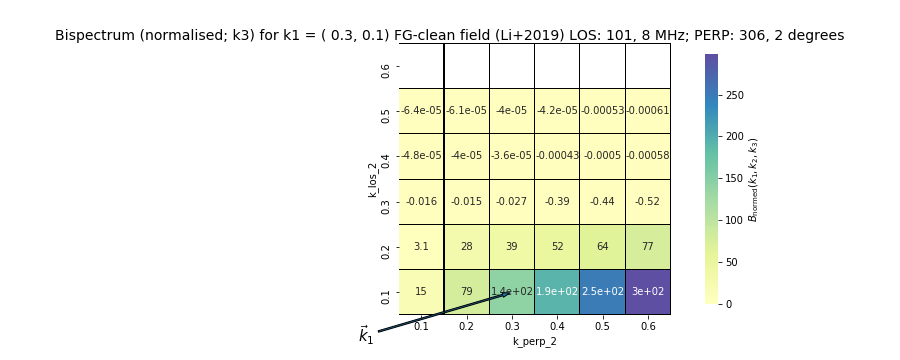} \\
    \includegraphics[trim=10.9cm 0.0cm 4.0cm 1.5cm, clip=true, scale=0.5]{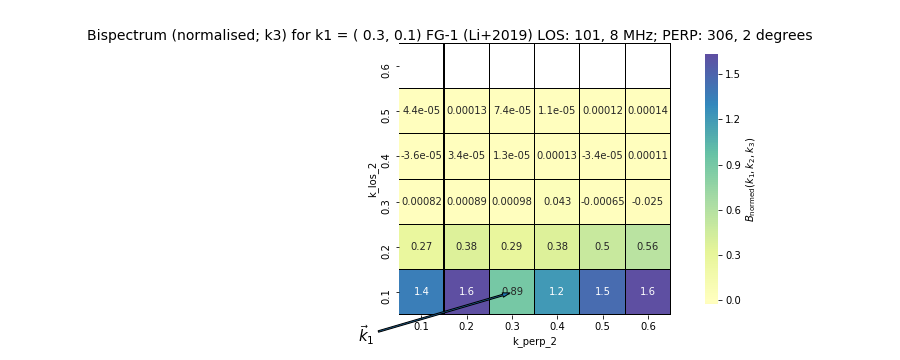} \\
  \end{array}$
  \caption{Normalised bispectrum for $(\kperp_1, \klos_1) = (0.3$ \invmpc, 0.1 \invmpc) from clean
  FG-only (field 1; top) and from the "observed" FG-only (field 1; bottom).
  We plot this as a function of $\kperp_2$ and
  $\klos_2$ so that each square corresponds to a different $\vect{k}$-triangle configuration.
  The simulation of observations suppresses the FG-only bispectrum, and qualitatively
  changes it from that of the clean field 1.}
  \label{fig:clean_vs_obs_pt1_pt3}
\end{figure}

\begin{figure}%[!hbtp]
\centering
  $\renewcommand{\arraystretch}{-0.75}
  \begin{array}{c}
    \includegraphics[trim=10.9cm 0.0cm 4.0cm 1.5cm, clip=true, scale=0.5]{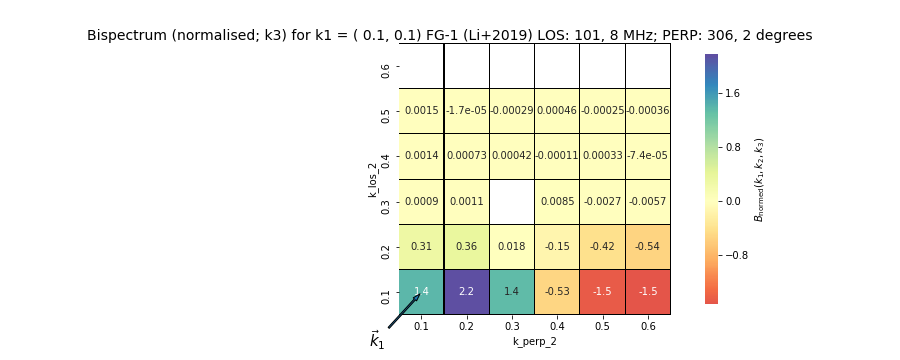} \\
  \end{array}$
  \caption{Normalised bispectrum for $(\kperp_1, \klos_1) = (0.1$ \invmpc, 0.1 \invmpc) from
  "observed" FG-only (field 1).
  This is the same field as the clean FG-only whose bispectrum is featured in the
  middle plot of Fig.~\ref{fig:kperp_pt1_clean}.
  We plot this as a function of $\kperp_2$ and
  $\klos_2$ so that each square corresponds to a different $\vect{k}$-triangle configuration.
  For this configuration set, the simulating of observations invert the sign on smaller scales
  $\kperp_2 > 0.3$ \invmpc and boost the bispectrum on larger scales $\kperp_2 < 0.3$ \invmpc.}
  \label{fig:obs_pt1_pt1}
\end{figure}

In this section we consider to what degree the instrumentals might alter the
bispectrum.
Fig.~\ref{fig:clean_vs_obs_pt1_pt3} shows the bispectrum from the clean FG-only (field 1; top)
and for the FG-only simulated observation (field 1; bottom) for the
$(\kperp_1, \klos_1) = (0.3$ \invmpc, 0.1 \invmpc) configuration set.
We note that the bispectrum is far stronger for this set of configurations than
it was for $(\kperp_1, \klos_1) = (0.1$ \invmpc, 0.1 \invmpc)
(see Fig.~\ref{fig:kperp_pt1_clean}) and $(\kperp_1, \klos_1) = (0.3$ \invmpc, 0.3 \invmpc) (see Fig.~\ref{fig:kperp_pt3_clean})
and, as with $(\kperp_1, \klos_1) = (0.1$ \invmpc, 0.1 \invmpc), is confined to $\klos_1<0.2$ \invmpc.
In such a case where the bispectrum is so strong, the imprint of this OSKAR simulated
instrumentals drastically suppresses the magnitude of the bispectrum
in such a way as to alter it qualitatively.
After simulating observational effects, the bispectrum for larger scales
($\kperp_2 < 0.3$ \invmpc)
is as strong as it is at $\kperp_2 > 0.3$ \invmpc,
whereas in the clean field it drops off monotonically from $\kperp_2 = 0.6$ \invmpc until it is two
($\klos_2 = 0.2$ \invmpc) to three ($\klos_2 = 0.1$ \invmpc)
orders of magnitude by $\kperp_2 = 0.1$ \invmpc.
For this configuration set, the positive sign of the bispectrum is maintained
after simulating observational effects.
However, when the foreground bispectrum is smaller, such
as is the case for $(\kperp_1, \klos_1) = (0.1$ \invmpc, 0.1 \invmpc), the bispectrum sign is altered for some configurations.
This can be seen by comparing Fig.~\ref{fig:obs_pt1_pt1} in which we show the bispectrum
from the "observed" field 1 to the bispectrum of the clean field 1
(see the middle plot of Fig.~\ref{fig:kperp_pt1_clean}).
What is clear from comparing with the clean bispectrum, is that the bispectrum is
actually boosted to have a larger positive bispectrum at $\kperp_2 < 0.3$ \invmpc in
the simulated observation than it does in the clean field 1.
It is clear from this that instrumental effects, even if we ignore complications
associated with things like calibration errors and ionospheric effects,
have the potential to seriously corrupt the bispectrum and would ultimately render
it useless for constraining the cosmological signal.
Although, such sensitivity to instrumental effects could be beneficial for refining
our data processing pipeline.

In the absence of $uv$ cutoff of 1000$\, \lambda$ (as performed in the imaging of the Li2019 datasets), the SKA imaging performance is exceptional, in both snapshot mode and with rotation synthesis. It is expected then that SKA images should be clean of sidelobes with little deconvolution required.
As such the findings in this section are very much tentative, and motivate more
detailed studies into the effects of instrumental effects on the bispectrum.

\begin{figure}%[!hbtp]
\centering
  $\renewcommand{\arraystretch}{-0.75}
  \begin{array}{c}
    \includegraphics[trim=10.9cm 0.0cm 4.0cm 1.5cm, clip=true, scale=0.5]{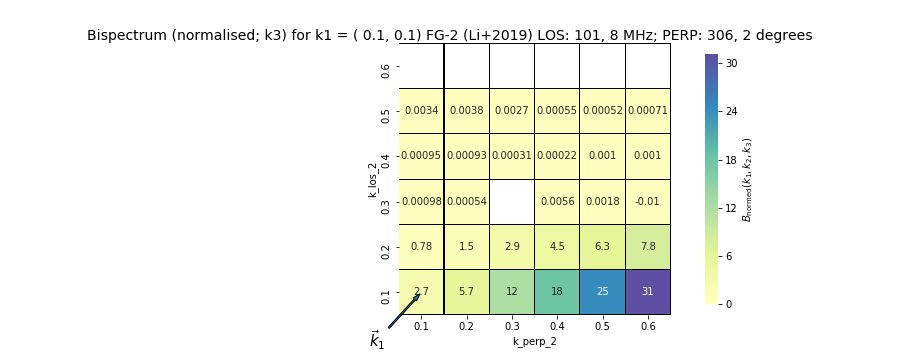} \\
  \end{array}$
  \caption{Normalised bispectrum for $(\kperp_1, \klos_1) = (0.1$ \invmpc, 0.1 \invmpc) from
  "observed" FG-only (field 2).
  We plot this as a function of $\kperp_2$ and
  $\klos_2$ so that each square corresponds to a different $\vect{k}$-triangle configuration.
  This differs substantially from that of "observed" field 1, with large amplitude and a
  bispectrum qualitatively closer to that of clean field 1.
  It is likely that the lack of concentrated FG sources in this field
  means it suffers from less corruption when observational effects are added.}
  \label{fig:obs2_pt1_pt1}
\end{figure}

As well as instrumental effects complicating our ability to interpret the 21cm bispectrum,
it is also prudent to consider the effects of sample variance on the bispectrum
of the foregrounds.
We therefore compare the bispectrum of "observed" FG-only (field 2; shown in Fig.~\ref{fig:obs2_pt1_pt1}) with
that of the "observed" FG-only field 1 (Fig.~\ref{fig:obs_pt1_pt1}) for $(\kperp_1, \klos_1) = (0.1$ \invmpc, 0.1 \invmpc).
For this configuration set, the differences between the two fields are major.
The "observed" bispectrum for field 2 is considerably larger for this configuration set;
even after suppression by instrumental effects it is still quite large and positive.
It is therefore either drastically larger before observational effects are simulated,
or the instrumental effects have less impact on the bispectrum of field 2.
We deem the latter the more likely since the foregrounds in field 2
are dominated by diffuse foregrounds and therefore contains less small-scale structure
that may be corrupted.
This intuition is further backed up by the fact that there are also configuration sets
for which the bispectrum of these two fields are quite similar, such as
$(\kperp_1, \klos_1) = (0.1$ \invmpc, 0.3 \invmpc) - not shown.
Naturally it would be preferable to compare the FG-only fields clean, however Li2019
do not include clean datasets for field 2.
Furthermore, it would be desirable to look at a slightly larger field of view,
as a 2$^\circ$ is a relatively conservative cropping choice for SKA.
We again defer detailed consideration of this point to future work.

%%%%%%%%%%%%%%%%%%%%%%%%%%%%%%%%%%%%%%%%%%%%%%%%%%%%%%%%%%%%%%%%%%%%%%%%%%%%%%%%
%%%%%%%%%%%%%%%%%%%%%%%%%%%%%%%%%%%%%%%%%%%%%%%%%%%%%%%%%%%%%%%%%%%%%%%%%%%%%%%%
\section{Prospects for recovering the bispectrum with foreground removal using GMCA} \label{sec:GMCA}
%%%%%%%%%%%%%%%%%%%%%%%%%%%%%%%%%%%%%%%%%%%%%%%%%%%%%%%%%%%%%%%%%%%%%%%%%%%%%%%%
%%%%%%%%%%%%%%%%%%%%%%%%%%%%%%%%%%%%%%%%%%%%%%%%%%%%%%%%%%%%%%%%%%%%%%%%%%%%%%%%

Given that we have seen that there is no clear EoR window for the bispectrum as there is
for the power spectrum, we take an initial look at the prospects at recovering the
signal using foreground removal.
We apply Generalized Morphological Component Analysis (GMCA) to both the clean
FG+EoR (field 1) to consider an ideal case,
as well as the "observed" FG+EoR (field 1).
GMCA exploits sparseness of signals in a particular basis, here a wavelet deconstruction,
in order to perform blind source separation (BSS). This BSS estimates a mixing matrix and
signal combination that maximises the sparseness of the signal.
This produces a recovered signal and a noise residual.
We refer the interested reader to \citealt{Bobin2007}, \citealt{Bobin2008}, and \citealt{Bobin2013}
for details of this algorithm.
Because the foregrounds are so many orders of magnitude greater than the EoR signal,
GMCA works to recover the foreground as the signal, leaving the EoR signal as part of the noise residuals.
Whilst GMCA technically performs blind source separation, it does require a little
guidance in terms of being told how many statistically-independent (linearly-combined)
components the signal, in our case the foregrounds, consists of.
This is ultimately a free parameter, if too small the algorithm will struggle to
accurately recover the signal,
and if too large, it will overfit and might also recover the EoR signal;
i.e. if it is not chosen correctly the algorithm will fail to separate the EoR
signal out as residuals.
For this work, we use 4 components
which were found by \citealt{Chapman2016} to be the optimal choice for recovering
the power spectrum from OSKAR-simulated LOFAR data.
We have experimented with a range of component numbers, and our initial findings
agree with those of \citealt{Chapman2016}.
However, we will perform a more thorough investigation of foreground removal in the context
of bispectrum recovery in a future study, including consideration
of other foreground removal methods than GMCA.

\begin{figure}%[!hbtp]
  \centering
    $\renewcommand{\arraystretch}{-0.75}
    \begin{array}{c}
      \includegraphics[trim=10.9cm 0.0cm 4.0cm 1.5cm, clip=true, scale=0.5]{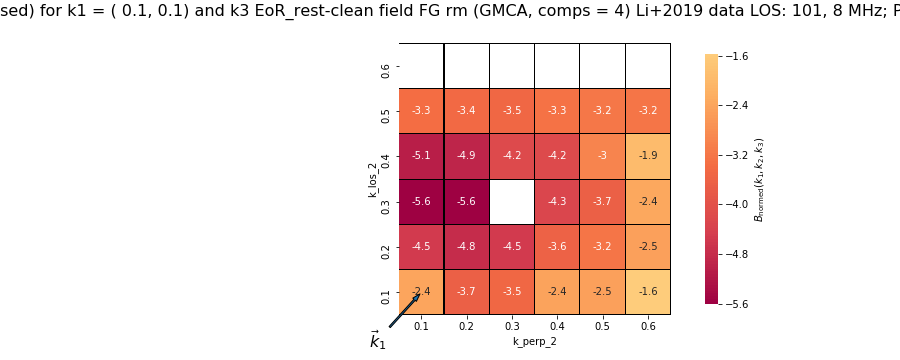}\\
      \includegraphics[trim=10.9cm 0.0cm 4.0cm 1.5cm, clip=true, scale=0.5]{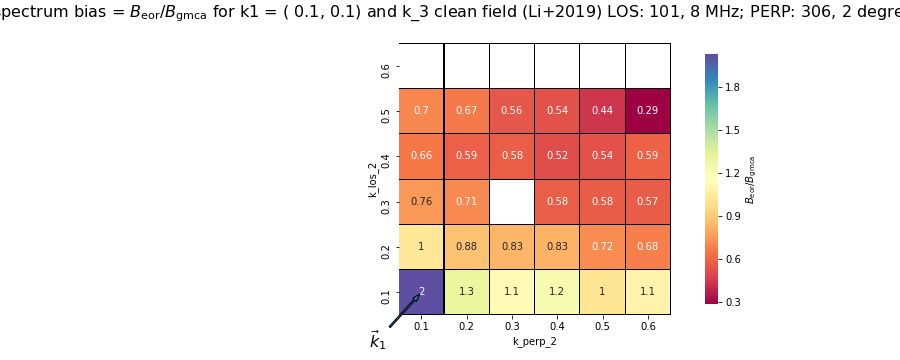}\\
    \end{array}$
  \caption{Normalised bispectrum from GMCA (4 components) residuals from clean
  EoR+FG (field 1; top) for $(\kperp_1, \klos_1) = (0.1$ \invmpc, 0.1 \invmpc).
  We plot this as a function of $\kperp_2$ and
  $\klos_2$ so that each square corresponds to a different $\vect{k}$-triangle configuration.
  In the absence of instrumental effects, GMCA fails to qualitatively recover
  the EoR-only bispectrum over the full configuration space,
  although the bispectrum of the residuals exhibits the correct order
  of magnitude and sign.
  The bottom plot shows the likeness ratio $B_{\mathrm{eor}}/B_{\mathrm{gmca}}$ of the
  GMCA-extracted bispectrum to the true bispectrum.
  We see there are certain configurations for which the recovery is good for which the
  likeness ratio is close to one.}
  \label{fig:gmca_clean_pt1}
\end{figure}

The top plot of Fig.~\ref{fig:gmca_clean_pt1} shows the $(\kperp_1, \klos_1) = (0.1$ \invmpc, 0.1 \invmpc)
bispectrum of the residuals after running GMCA on clean EoR+FG (field 1) which, if
working perfectly, would reproduce the clean EoR-only bispectrum.
In the bottom plot of Fig.~\ref{fig:gmca_clean_pt1} we show the \textbf{likeness ratio}
$B_{\mathrm{eor}}/B_{\mathrm{gmca}}$ of the bispectrum from clean EoR-only (field 1)
to that of the GMCA residuals.
Even in the absence of instrumental effects, GMCA does not qualitatively recover
the bispectrum very well, reaching its most negative at $\klos_2 = 0.3$ \invmpc, $\kperp_2 < 0.3$ \invmpc
rather than at $\klos_2 = \kperp_2 = 0.1$ \invmpc as seen in the top plot of Fig.~\ref{fig:kperp_pt1_clean}.
However, it certainly does recover the correct sign and order of magnitude which
would still be useful information to have, so long as we can characterise the
error and/or bias from foreground removal.
There are also certain configurations for which GMCA performs quite well at recovering
the bispectrum for, for example $\klos_2 = 0.1$ \invmpc and 0.3 \invmpc $\le \kperp_2 \le 0.6$ \invmpc
where we see the likeness ratio of the bottom plot of Fig.~\ref{fig:gmca_clean_pt1}
is close to one.

\begin{figure}%[!hbtp]
  \centering
    $\renewcommand{\arraystretch}{-0.75}
    \begin{array}{c}
      \includegraphics[trim=10.9cm 0.0cm 4.0cm 1.5cm, clip=true, scale=0.5]{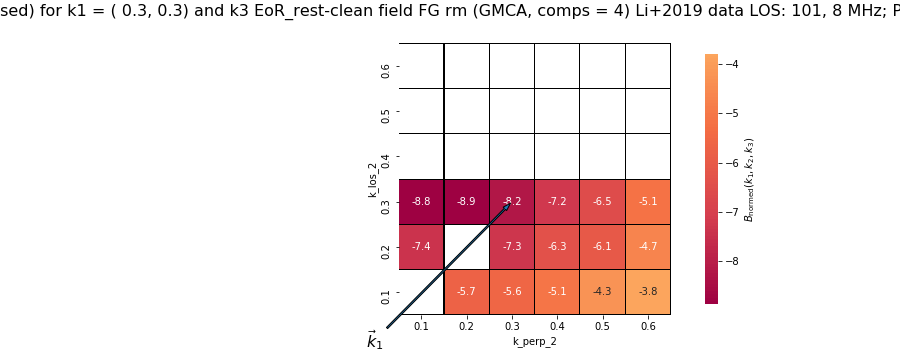}\\
      \includegraphics[trim=10.9cm 0.0cm 4.0cm 1.5cm, clip=true, scale=0.5]{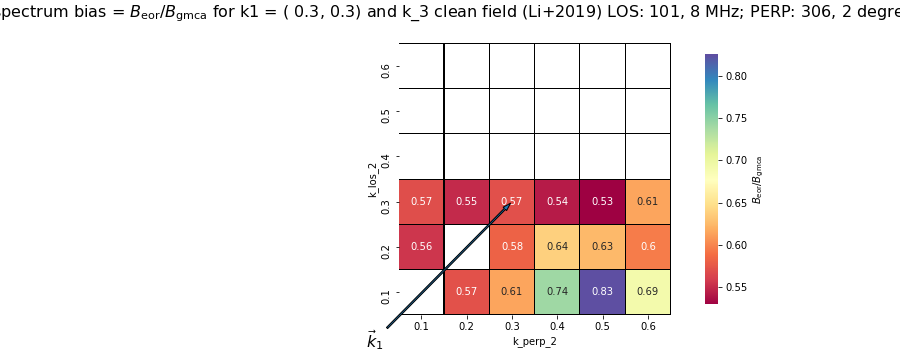}\\
    \end{array}$
  \caption{Normalised bispectrum from GMCA (4 components) residuals from the clean
  EoR+FG (field 1; top) for $(\kperp_1, \klos_1) = (0.3$ \invmpc, 0.3 \invmpc). We plot this as a function of $\kperp_2$ and
  $\klos_2$ so that each square corresponds to a different $\vect{k}$-triangle configuration.
  The bottom plot shows the likeness ratio $B_{\mathrm{eor}}/B_{\mathrm{gmca}}$ of the GMCA-extracted
  bispectrum to the true bispectrum.
  For this configuration set, the qualitative recovery of the bispectrum is
  good, with a bias in amplitude producing a likeness ratio that is consistently $\sim 0.6$.}
  \label{fig:gmca_clean_pt3}
\end{figure}

We show the $(\kperp_1, \klos_1) = (0.3$ \invmpc, 0.3 \invmpc) bispectrum of the GMCA residuals
and the likeness ratio in the top and bottom plots of
Fig.~\ref{fig:gmca_clean_pt3} respectively.
On such smaller scales, GMCA does a better job of recovering the qualitative
behaviour of the EoR bispectrum over configuration space,
with its bispectrum consistently
biased so that $B_{\mathrm{eor}}/B_{\mathrm{gmca}} \sim 0.6$ for all configurations.

\begin{figure}%[!hbtp]
  \centering
    $\renewcommand{\arraystretch}{-0.75}
    \begin{array}{c}
      \includegraphics[trim=10.9cm 0.0cm 4.0cm 1.5cm, clip=true, scale=0.5]{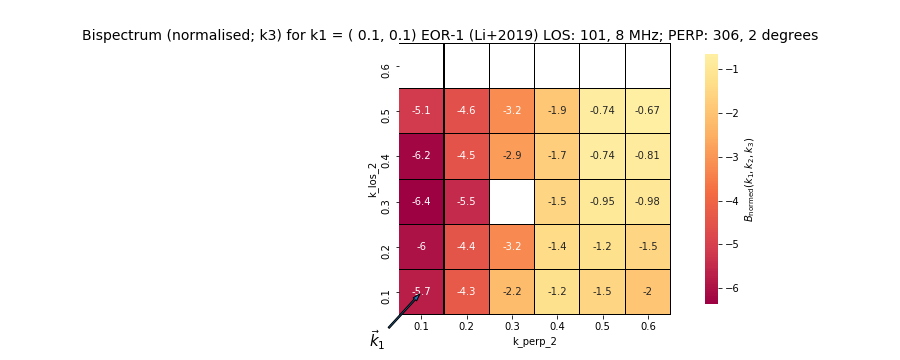}\\
      \includegraphics[trim=10.9cm 0.0cm 4.0cm 1.5cm, clip=true, scale=0.5]{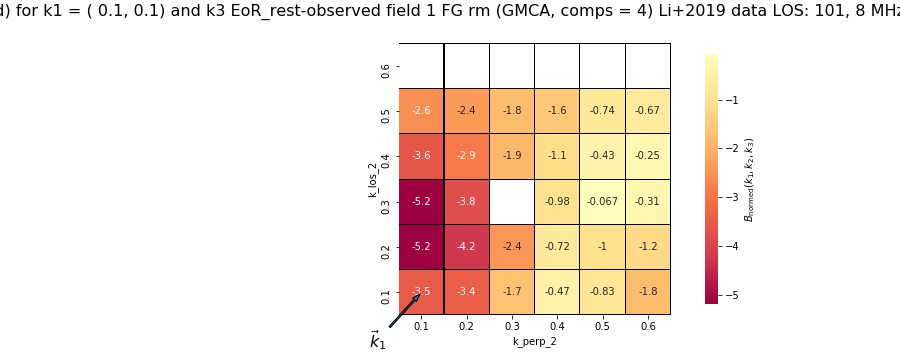}\\
    \end{array}$
  \caption{Normalised bispectrum from "observed" EoR-only
  (field 1; top) and the normalised bispectrum from the corresponding GMCA
  (4 components) residuals (bottom) for the $(\kperp_1, \klos_1) = (0.1$ \invmpc, 0.1 \invmpc)
  configuration set. We plot this as a function of $\kperp_2$ and
  $\klos_2$ so that each square corresponds to a different $\vect{k}$-triangle configuration.
  GMCA fails to recover the qualitative evolution of the bispectrum over configuration
  space, but the bispectrum of the GMCA residuals exhibit the correct sign and order of magnitude.}
  \label{fig:gmca_obs_pt1}
\end{figure}

Although we have seen that the bispectrum of the foregrounds
are seriously corrupted by the simulation of observational effects,
it is still interesting to take a first look at how GMCA performs at recovering
the bispectrum from the "observed" Li2019 datasets.
We find that in contrast to when applied to the clean field, GMCA does worse at recovering the bispectrum
for $(\kperp_1, \klos_1) = (0.3$ \invmpc, 0.3 \invmpc) than it does for $(\kperp_1, \klos_1) = (0.1$ \invmpc, 0.1 \invmpc),
failing completely in sign and amplitude of the bispectrum beyond $\kperp_2>0.3$ \invmpc.
It is in this regime that the foreground bispectrum is very strong,
so this is perhaps unsurprising.
Its performance is comparable to that of $(\kperp_1, \klos_1) = (0.1$ \invmpc, 0.1 \invmpc) for
$\kperp_2\le 0.3$ \invmpc and so
we will focus our discussion on the performance of GMCA for the $(\kperp_1, \klos_1) = (0.1$ \invmpc, 0.1 \invmpc)
configuration set.

In the top plot of Fig.~\ref{fig:gmca_obs_pt1} we show the bispectrum for "observed" EoR-only (field 1)
for the $(\kperp_1, \klos_1) = (0.1$ \invmpc, 0.1 \invmpc) configuration set.
The bottom plot of Fig.~\ref{fig:gmca_obs_pt1} shows the $(\kperp_1, \klos_1) = (0.1$ \invmpc, 0.1 \invmpc)
bispectrum from the residuals after running GMCA on "observed" EoR+FG (field 1).
We can see from this plot that promisingly GMCA, even in the presence of instrumental effects,
successfully recovers the correct sign and order of magnitude as in "observed" EoR-only (field 1).
As with the clean field, the GMCA-residual's bispectrum does not
qualitatively agree with the "observed" EoR-only bispectrum.

\begin{figure}%[!hbtp]
  \centering
    $\renewcommand{\arraystretch}{-0.75}
    \begin{array}{c}
      \includegraphics[trim=10.9cm 0.0cm 4.0cm 1.5cm, clip=true, scale=0.5]{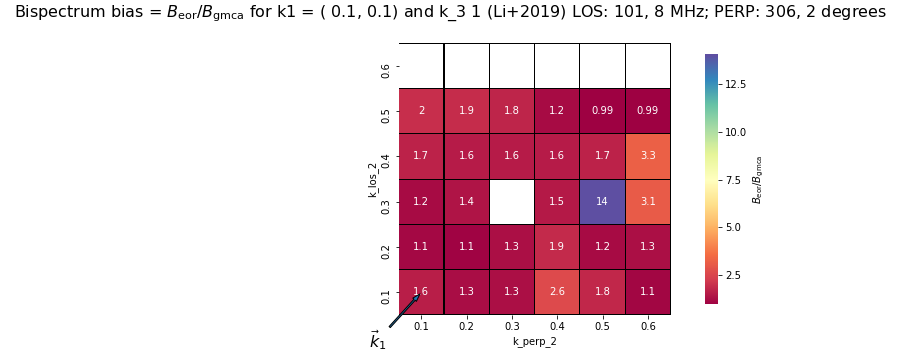}\\
      \includegraphics[trim=10.9cm 0.0cm 4.0cm 1.5cm, clip=true, scale=0.5]{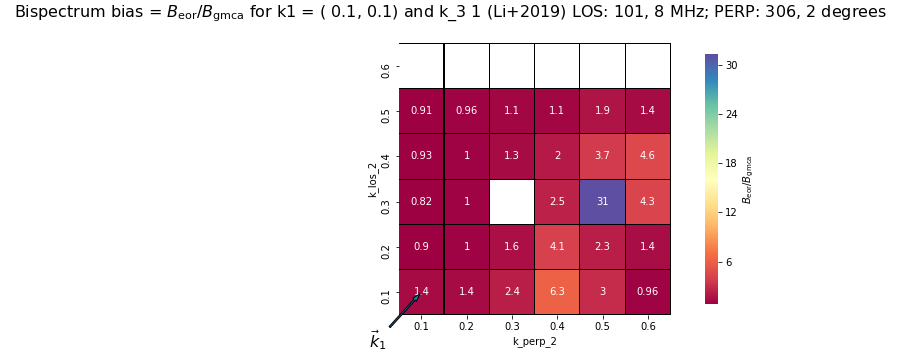}\\
    \end{array}$
  \caption{Likeness ratio $B_{\mathrm{eor}}/B_{\mathrm{gmca}}$ of the normalised
  bispectrum of "observed" EoR-only (field 1) compared to that of the corresponding
  GMCA (4 components) residuals from EoR+FG (top).
  The bottom plot shows the likeness ratio of the bispectrum of clean EoR-only (field 1) to that of the same GMCA residuals.
  Both for the $(\kperp_1, \klos_1) = (0.1$ \invmpc, 0.1 \invmpc) configuration set.
  We plot this as a function of $\kperp_2$ and
  $\klos_2$ so that each square corresponds to a different $\vect{k}$-triangle configuration.
  GMCA seems to perform better at recovering the clean EoR-only bispectrum from the "observed" field 1
  for $\kperp_2\le 0.3$ \invmpc, indicating that instrumental effect here modulate
  the foreground field so that it is better described by four independent components.}
  \label{fig:corr_obs_pt1}
\end{figure}

In the top plot of Fig.~\ref{fig:corr_obs_pt1} we show the likeness ratio
$B_{\mathrm{eor}}/B_{\mathrm{gmca}}$, here comparing the bispectrum from GMCA
residuals from the "observed" EoR+FG with the "observed" EoR-only (both for field 1).
Whilst the bispectrum of the "observed" EoR-only fields is not technically the quantity
we are aiming to recover, the particular observation simulation we have
analysed is potentially not typical, so it is useful to understand how well
it can recover the "observed" EoR-only bispectrum as well as that of clean EoR-only.
It is also in principle possible to forward model the observation pipeline for parameter estimation purposes (although it remains to be seen how practical this is in reality).
As with the $(\kperp_1, \klos_1) = (0.3$ \invmpc, 0.3 \invmpc) configuration set, GMCA struggles
for $\kperp_2>0.3$ \invmpc and $\klos_2>0.3$ \invmpc.
This is interesting as the foreground bispectrum is negligible at $\klos_2>0.3$ \invmpc
even in the "observed" field 1.

In the bottom plot of \ref{fig:corr_obs_pt1} we show $B_{\mathrm{eor}}/B_{\mathrm{gmca}}$ with $B_{\mathrm{eor}}$ being from
clean EoR-only (field 1) and $B_{\mathrm{gmca}}$ from the residuals from the "observed" field 1.
Interestingly, for the GMCA residuals from the "observed" field 1 seem to exhibit a bispectrum
that is closer to the true bispectrum for $\kperp_2\le 0.3$ \invmpc than the residuals from
the clean field.
It is possible then, that the instrumental corruption somehow aids the GMCA source separation.
This intuitively makes sense; for example, instrumental smoothing could simplify the foregrounds so that they are better described as 4 statistically-independent components,
i.e. the complexity of the foregrounds might be reduced by the inclusion of instrumental effects.
It is of course entirely possible, that this apparent improvement in performance
is by chance. This point therefore warrants further investigation, which we will
address in a future study.

It would be remiss to judge our ability to recover the EoR bispectrum from the "observed" dataset discussed here without considering the associated GMCA recovery of the power spectrum.
We do not include plots of the 2D power spectrum measured from GMCA residuals in the main text. We instead include them in Appendix \ref{appendix:gmcaPS} as they are interesting in their own right, but are not essential to our discussion.
GMCA is excellent at recovering the "observed" EoR power spectrum outside of the wedge.
So assuming we can effectively forward model the observation, the power spectrum is less impacted by foreground residuals.
However, the 2D power spectrum of the GMCA residuals do not compare so well with that of the clean EoR signal.
The likeness ratio for $P_{\mathrm{GMCA}}/P_{\mathrm{EoR}}$ is never 1 (which it is with the bispectrum for mant configurations) and is only less than 1.1 for $\kperp<0.2$.

Whilst it appears that GMCA cannot accurately recover the bispectrum (or the power spectrum for that matter) from this dataset,
there are other foreground-removal algorithms such as the Gaussian processes method of \citealt{Mertens2017a}.
There is also the promising option to use a convolutional denoising autoencoder to learn the features of the signal and therefore separate the signal from the foregrounds,
as done by \citealt{Li2019}.
It is quite possible that one of those or another foreground-removal
algorithm will do a better job than GMCA at recovering the EoR bispectrum.
We will also consider this question further in a future study.

The foreground simulations analysed here assume that point sources with a flux greater than 10 mJy have been removed. We have analysed the point-source 21cm bispectrum as predicted by \citealt{Trott2019}, and find that the bispectrum will be boosted in magnitude (relative to that of the Li2019 foregrounds) if point source are not so effectively removed. For example, if we assume that only point sources with a flux greater than 50 mJy can be removed, then the maximum normalised bispectrum can be as high as  $10^{4}$ for $(\kperp_1, \klos_1) = (0.1$ \invmpc, 0.1 \invmpc) and  $10^{56}$ for $(\kperp_1, \klos_1) = (0.3$ \invmpc, 0.3 \invmpc).
It is therefore important that alongside developing an understanding of the impact of foreground residuals on the 21cm bispectrum, that we also understand of the impact of point-source removal residuals.

%%%%%%%%%%%%%%%%%%%%%%%%%%%%%%%%%%%%%%%%%%%%%%%%%%%%%%%%%%%%%%%%%%%%%%%%%%%%%%%%
%%%%%%%%%%%%%%%%%%%%%%%%%%%%%%%%%%%%%%%%%%%%%%%%%%%%%%%%%%%%%%%%%%%%%%%%%%%%%%%%
\section{Conclusion} \label{sec:Conc}
%%%%%%%%%%%%%%%%%%%%%%%%%%%%%%%%%%%%%%%%%%%%%%%%%%%%%%%%%%%%%%%%%%%%%%%%%%%%%%%%
%%%%%%%%%%%%%%%%%%%%%%%%%%%%%%%%%%%%%%%%%%%%%%%%%%%%%%%%%%%%%%%%%%%%%%%%%%%%%%%%
In this paper we have measured the bispectrum from accurate simulations of 21cm Foregrounds,
a typical Epoch of Reionisation simulation, and their combination.
We have also measured the bispectrum from these datasets after having been passed
through an SKA telescope observation pipeline (consisting of OSKAR + \wsclean).

Through our analysis, we have established that there is not a clean EoR window
for the bispectrum, this means that (unlike with the power spectrum) foreground
avoidance is not obviously a viable approach to constrain the 21cm EoR bispectrum.
The presence of EoR structure does however alter the bispectrum of the combined
field from that of the foreground-only field on smaller scales.
However, it does so through complex modulation of its structures with that of the foregrounds,
making it very difficult to interpret.

This is further complicated by our findings that simulating instrumental effects
can substantially alter the bispectrum, suppressing its amplitude on some scales,
introducing additional bispectrum amplitude on other scales, even inverting the sign
of the bispectrum on certain scales.
All of these effects can mimic the modulation effects of the EoR structure in the EoR+FG
bispectrum.
This warrants further study to understand the subtleties of instrumental effects
on the 21cm bispectrum. Even if it transpires these studies find the bispectrum cannot be
used to constrain the underlying EoR signal, it may well be that the bispectrum
will instead be useful for refining our processing of observed data more precisely.

Given the absence of an EoR window for the bispectrum, we take an initial look
at whether foreground removal can accurately recover the non-Gaussianities of
the EoR signal.
We assume four independent components to describe and fit the foreground
signal, which under perfect performance should return the EoR signal in its noise residuals.
The bispectrum of the GMCA residuals have the same sign as that of the EoR signal.
Their amplitude is also of the correct order of magnitude.
However, the qualitative evolution of the residual's bispectrum across configuration
space is not the same, apart from on smaller $k_1 = (\kperp_1, \klos_1)$ scales for the clean signal, i.e. in the
absence of simulated instrumentals.
When applied to the simulated observations, on larger scales ($\kperp \le 0.3$ \invmpc), the bispectrum of GMCA residuals
are closer to those of the clean EoR signal than the GMCA residuals from the clean EoR+FG field.
This is likely because the instrumental effects simplify the foreground signal
so that it is more cleanly described as four linearly-independent components, as assumed
by GMCA.
However, on smaller scales ($\kperp > 0.3$ \invmpc), GMCA fails entirely to recover
the bispectrum of the EoR.
It is worth noting that for this dataset, the true EoR bipectrum is more accurately recovered by GMCA than is the true 2D power spectrum (although, the 2D power spectrum of the observed EoR is recovered well outside of the wedge by GMCA).
These findings are encouraging and motivate more detailed studies to establish
how effective GMCA and other foreground removal algorithms are at the level of the bispectrum.

There is an important caveat to this study that is worth bearing in mind and should be
addressed in future studies. For the simulated observations analysed here there is a corruption to the signal that presents as a thin band of excess power at $\kperp \sim 0.7$ \invmpc and
a suppression of power beyond. This is caused by a $\lambda$ cut at the imaging stage. Although we have omitted all modes where $\kperp > 0.6$ to avoid this corruption,
it remains unclear how much this is causing the corruption of the bispectrum by instrumentals discussed in this paper.

Despite this caveat, the conclusions in this paper are on solid ground and motivate
further work to understand exactly how we might use the bispectrum in practice.
The rewards of such studies are a potentially more robust and accurate understanding
of the first stars and galaxies, and/or an important tool for improving our processing
of 21cm observations.

%%%%%%%%%%%%%%%%%%%%%%%%%%%%%%%%%%%%%%%%%%%%%%%%%%%%%%%%%%%%%%%%%%%%%%%%%%%%%%%%
%%%%%%%%%%%%%%%%%%%%%%%%%%%%%%%%%%%%%%%%%%%%%%%%%%%%%%%%%%%%%%%%%%%%%%%%%%%%%%%%
\section*{Acknowledgements}
%%%%%%%%%%%%%%%%%%%%%%%%%%%%%%%%%%%%%%%%%%%%%%%%%%%%%%%%%%%%%%%%%%%%%%%%%%%%%%%%
%%%%%%%%%%%%%%%%%%%%%%%%%%%%%%%%%%%%%%%%%%%%%%%%%%%%%%%%%%%%%%%%%%%%%%%%%%%%%%%%
CAW's research is supported by a UK Research and Innovation Future Leaders Fellowship, grant number MR/S016066/1.
However, the research presented in this paper was carried out with financial support from the European Research
Council under ERC grant number 638743-FIRSTDAWN (held by
Jonathan Pritchard). CAW also thanks Weitan Li for making his datasets and analysis
pipelines publicly available and for providing such good documentation.
IH acknowledges studentship funding from the Royal Society Dorothy Hodgkin Fellowship (held by Emma Chapman). This research was partly supported by the Australian Research Council Centre of Excellence for All Sky Astrophysics in 3 Dimensions (ASTRO 3D), through project number CE170100013.
CMT acknowledges financial support from an ARC Future Fellowship under grant FT180100321.

%%%%%%%%%%%%%%%%%%%%%%%%%%%%%%%%%%%%%%%%%%%%%%%%%%%%%%%%%%%%%%%%%%%%%%%%%%%%%%%%
%%%%%%%%%%%%%%%%%%%%%%%%%%%%%%%%%%%%%%%%%%%%%%%%%%%%%%%%%%%%%%%%%%%%%%%%%%%%%%%%
\bibliographystyle{mn2e}
%\bibliography{library}

\begin{thebibliography}{}
 \providecommand{\href}[2]{#2}
  \providecommand{\doi}[1]{\href{http://dx.doi.org/#1}{doi:#1}}
  \providecommand{\eprint}[1]{\href{http://arxiv.org/abs/#1}{arXiv:#1}}

\bibitem[\protect\citeauthoryear{Beardsley et~al.,}{Beardsley
  et~al.}{2016}]{Beardsley2016a}
Beardsley A.~P.  et~al., 2016, Astrophys. J., 833, 21, \eprint{1608.06281},
  \doi{10.3847/1538-4357/833/1/102}

\bibitem[\protect\citeauthoryear{Bobin, Moudden, Starck, Fadili \&
  Aghanim}{Bobin et~al.}{2008a}]{Bobin2008}
Bobin J.,  Moudden Y.,  Starck J.~L.,  Fadili J.,    Aghanim N.,  2008a, Stat.
  Methodol., 5, 307, \eprint{0712.0588}, \doi{10.1016/j.stamet.2007.10.003}

\bibitem[\protect\citeauthoryear{Bobin, Starck, Fadili \& Moudden}{Bobin
  et~al.}{2007b}]{Bobin2007}
Bobin J.,  Starck J.~L.,  Fadili J.,    Moudden Y.,  2007b, IEEE Trans. Image
  Process., 16, 2662, \doi{10.1109/TIP.2007.906256}

\bibitem[\protect\citeauthoryear{Bobin, Starck, Sureau \& Basak}{Bobin
  et~al.}{2013c}]{Bobin2013}
Bobin J.,  Starck J.~L.,  Sureau F.,    Basak S.,  2013c, A{\&}A, 550,
  \eprint{1206.1773}, \doi{10.1051/0004-6361/201219781}

\bibitem[\protect\citeauthoryear{Bowman, Rogers, Monsalve, Mozdzen \&
  Mahesh}{Bowman et~al.}{2018}]{Bowman2018a}
Bowman J.~D.,  Rogers A. E.~E.,  Monsalve R.~A.,  Mozdzen T.~J.,    Mahesh N.,
  2018, Nature, 555, 67, \doi{10.1038/nature25792}

\bibitem[\protect\citeauthoryear{Bradley, Tauscher, Rapetti \& Burns}{Bradley
  et~al.}{2019}]{Bradley2019}
Bradley R.~F.,  Tauscher K.,  Rapetti D.,    Burns J.~O.,  2019, Astrophys. J.,
  874, 153, \eprint{1810.09015}, \doi{10.3847/1538-4357/ab0d8b}

\bibitem[\protect\citeauthoryear{Braun, Bonaldi, Bourke, Keane \& Wagg}{Braun
  et~al.}{2019}]{Braun2019}
Braun R.,  Bonaldi A.,  Bourke T.,  Keane E.,    Wagg J.,  2019, Sq. Km. Array
  Memo, \eprint{1912.12699}

\bibitem[\protect\citeauthoryear{Chapman, Zaroubi, Abdalla, Dulwich,
  Jeli{\'{c}} \& Mort}{Chapman et~al.}{2016}]{Chapman2016}
Chapman E.,  Zaroubi S.,  Abdalla F.~B.,  Dulwich F.,  Jeli{\'{c}} V.,    Mort
  B.,  2016, MNRAS, 458, 2928, \doi{10.1093/mnras/stw161}

\bibitem[\protect\citeauthoryear{Datta, Bowman \& Carilli}{Datta
  et~al.}{2010}]{Datta2010}
Datta A.,  Bowman J.~D.,    Carilli C.~L.,  2010, Astrophys. J., 724, 526,
  \eprint{1005.4071}

\bibitem[\protect\citeauthoryear{{Di Matteo}, Ciardi \& Miniati}{{Di Matteo}
    et~al.}{2004}]{DiMatteo2004}
  {Di Matteo} T.,  Ciardi B.,    Miniati F.,  2004, MNRAS, 355, 1053,
    \doi{10.1111/j.1365-2966.2004.08443.x}

\bibitem[\protect\citeauthoryear{Finkbeiner}{Finkbeiner}{2003}]{Finkbeiner2003}
Finkbeiner D.~P.,  2003, Astrophys. J. Suppl. Ser., 146, 407, \eprint{0301558},
  \doi{10.1086/374411}

\bibitem[\protect\citeauthoryear{Gehlot et~al.,}{Gehlot
  et~al.}{2019}]{Gehlot2018}
Gehlot B.~K.  et~al., 2019, MNRAS, 488, 4271, \eprint{1809.06661},
  \doi{10.1093/mnras/stz1937}

\bibitem[\protect\citeauthoryear{Gleser, Nusser \& Benson}{Gleser
  et~al.}{2008}]{Gleser2008}
Gleser L.,  Nusser A.,    Benson A.~J.,  2008, MNRAS, 391, 383,
  \eprint{0712.0497v2}, \doi{10.1111/j.1365-2966.2008.13897.x}

\bibitem[\protect\citeauthoryear{Greig \& Mesinger}{Greig \&
  Mesinger}{2017}]{Greig2017}
Greig B.,  Mesinger A.,  2017, Mon. Not. R. Astronimcal Soc.,
  \eprint{1705.03471}

\bibitem[\protect\citeauthoryear{Haslam}{Haslam}{1983}]{Haslam1983}
Haslam C. G.~T.,  1983, Obs., 103, 133

\bibitem[\protect\citeauthoryear{Hazelton, Morales \& Sullivan}{Hazelton
  et~al.}{2013}]{Hazelton2013}
Hazelton B.~J.,  Morales M.~F.,    Sullivan I.~S.,  2013, Astrophys. J., 770,
  156, \doi{10.1088/0004-637X/770/2/156}

\bibitem[\protect\citeauthoryear{Hills, Kulkarni, Meerburg \& Puchwein}{Hills
  et~al.}{2018}]{Hills2018}
Hills R.,  Kulkarni G.,  Meerburg P.~D.,    Puchwein E.,  2018, Nature, 564,
  32, \eprint{1805.01421}

\bibitem[\protect\citeauthoryear{Hutter, Watkinson, Seiler, Dayal, Sinha \&
  Croton}{Hutter et~al.}{2019}]{Hutter2019}
Hutter A.,  Watkinson C.~A.,  Seiler J.,  Dayal P.,  Sinha M.,    Croton D.~J.,
   2019, MNRAS, \doi{10.1093/mnras/stz3139}

\bibitem[\protect\citeauthoryear{Kolopanis et~al.,}{Kolopanis
  et~al.}{2019}]{Kolopanis2019}
Kolopanis M.  et~al., 2019, Astrophys. J., 883, 133, \eprint{1909.02085},
  \doi{10.3847/1538-4357/ab3e3a}

\bibitem[\protect\citeauthoryear{Lewis}{Lewis}{2011}]{Lewis2011}
Lewis A.,  2011, J. Cosmol. Astropart. Phys., 10, 1475, \eprint{1107.5431},
  \doi{10.1088/1475-7516/2011/10/026}

\bibitem[\protect\citeauthoryear{Li et~al.,}{Li et~al.}{2019a}]{Li2019}
Li W.  et~al., 2019a, MNRAS, 485, 2628, \eprint{1902.09278},
  \doi{10.1093/mnras/stz582}

\bibitem[\protect\citeauthoryear{Li et~al.,}{Li et~al.}{2019b}]{Li2019b}
Li W.  et~al., 2019b, Astrophys. J., 879, 104, \eprint{1905.05399},
  \doi{10.3847/1538-4357/ab21bc}

\bibitem[\protect\citeauthoryear{Li et~al.,}{Li et~al.}{2019c}]{Li2019a}
Li W.  et~al., 2019c, Astrophys. J., 887, 141, \doi{10.3847/1538-4357/ab55e4}

\bibitem[\protect\citeauthoryear{Liu \& Tegmark}{Liu \&
  Tegmark}{2012}]{Liu2012}
Liu A.,  Tegmark M.,  2012, MNRAS, 419, 3491,
  \doi{10.1111/j.1365-2966.2011.19989.x}

\bibitem[\protect\citeauthoryear{Liu, Parsons \& Trott}{Liu
  et~al.}{2014}]{Liu2014c}
Liu A.,  Parsons A.~R.,    Trott C.~M.,  2014, Phys. Rev. D, 90, 023018,
  \doi{10.1103/PhysRevD.90.023018}

\bibitem[\protect\citeauthoryear{Loeb \& Furlanetto}{Loeb \&
  Furlanetto}{2013}]{Loeb2013}
Loeb A.,  Furlanetto S.~R.,  2013, {The First Galaxies in the Universe}.
Princeton University Press

\bibitem[\protect\citeauthoryear{Majumdar, Pritchard, Mondal, Watkinson,
  Bharadwaj \& Mellema}{Majumdar et~al.}{2017}]{Majumdar2017}
Majumdar S.,  Pritchard J.~R.,  Mondal R.,  Watkinson C.~A.,  Bharadwaj S.,
  Mellema G.,  2017, MNRAS, 476, 4007, \eprint{1708.08458},
  \doi{10.1093/mnras/sty535}

\bibitem[\protect\citeauthoryear{Mertens, Ghosh \& Koopmans}{Mertens
  et~al.}{2017}]{Mertens2017a}
Mertens F.~G.,  Ghosh A.,    Koopmans L. V.~E.,  2017, MNRAS,
  \eprint{1711.10834}, \doi{10.1093/mnras/sty1207}

\bibitem[\protect\citeauthoryear{Mesinger, Furlanetto \& Cen}{Mesinger
  et~al.}{2011}]{Signal2010}
Mesinger A.,  Furlanetto S.~R.,    Cen R.,  2011, MNRAS, 411, 955,
  \eprint{1003.3878v1}

\bibitem[\protect\citeauthoryear{Morales \& Hewitt}{Morales \&
  Hewitt}{2004}]{Morales2004}
Morales M.~F.,  Hewitt J.,  2004, Astrophys. J., 615, 7, \eprint{0312437},
  \doi{10.1086/424437}

\bibitem[\protect\citeauthoryear{Morales, Hazelton, Sullivan \&
  Beardsley}{Morales et~al.}{2012}]{Morales2012}
Morales M.~F.,  Hazelton B.,  Sullivan I.,    Beardsley A.,  2012, Astrophys.
  J., 752, 137, \doi{10.1088/0004-637X/752/2/137}

\bibitem[\protect\citeauthoryear{Murray, Trott \& Jordan}{Murray
  et~al.}{2017}]{Murray2017}
Murray S.~G.,  Trott C.~M.,    Jordan C.~H.,  2017, \eprint{1706.10033}

\bibitem[\protect\citeauthoryear{Offringa et~al.,}{Offringa
  et~al.}{2014}]{Offringa2014}
Offringa A.~R.  et~al., 2014, MNRAS, 444, 606, \doi{10.1093/mnras/stu1368}

\bibitem[\protect\citeauthoryear{Park, Mesinger, Greig \& Gillet}{Park
  et~al.}{2018}]{Park2018}
Park J.,  Mesinger A.,  Greig B.,    Gillet N.,  2018, Mon. Not. R. Astron.
  Soc. Vol. 484, Issue 1, p.933-949, 484, 933, \eprint{1809.08995},
  \doi{10.1093/mnras/stz032}

\bibitem[\protect\citeauthoryear{Patil et~al.,}{Patil et~al.}{2017}]{Patil2017}
Patil A.~H.  et~al., 2017, Astrophys. J., 838, 65, \eprint{1702.08679},
  \doi{10.3847/1538-4357/aa63e7}

\bibitem[\protect\citeauthoryear{Shaver, Windhorst, Madau \& de Bruyn}{Shaver
  et~al.}{1999}]{Shaver1999b}
Shaver P.~A.,  Windhorst R.~A.,  Madau P.,    de Bruyn A.~G.,  1999, A{\&}A,
  345, 380

\bibitem[\protect\citeauthoryear{Shimabukuro, Yoshiura, Takahashi, Yokoyama \&
  Ichiki}{Shimabukuro et~al.}{2016}]{Shimabukuro2016a}
Shimabukuro H.,  Yoshiura S.,  Takahashi K.,  Yokoyama S.,    Ichiki K.,  2016,
  MNRAS, 468, 1542, \eprint{1608.00372}, \doi{10.1093/mnras/stx530}

\bibitem[\protect\citeauthoryear{Sims \& Pober}{Sims \& Pober}{2020}]{Sims2020}
Sims P.~H.,  Pober J.~C.,  2020, MNRAS, 492, 22, \doi{10.1093/mnras/stz3388}

\bibitem[\protect\citeauthoryear{Spinelli, Bernardi \& Santos}{Spinelli
  et~al.}{2018}]{Spinelli2018}
Spinelli M.,  Bernardi G.,    Santos M.~G.,  2018, MNRAS, 479, 275,
  \eprint{1802.03060}, \doi{10.1093/mnras/sty1457}

\bibitem[\protect\citeauthoryear{Thyagarajan et~al.,}{Thyagarajan
  et~al.}{2013}]{Thyagarajan2013}
Thyagarajan N.  et~al., 2013, Astrophys. J., 776, 6,
  \doi{10.1088/0004-637X/776/1/6}

\bibitem[\protect\citeauthoryear{Trott et~al.,}{Trott et~al.}{2019}]{Trott2019}
Trott C.~M.  et~al., 2019, Publ. Astron. Soc. Aust., 36, e023,
  \eprint{1905.07161}, \doi{10.1017/pasa.2019.15}

\bibitem[\protect\citeauthoryear{Vedantham, {Udaya Shankar} \&
  Subrahmanyan}{Vedantham et~al.}{2012}]{Vedantham2012}
Vedantham H.,  {Udaya Shankar} N.,    Subrahmanyan R.,  2012, Astrophys. J.,
  745, 176, \doi{10.1088/0004-637X/745/2/176}

\bibitem[\protect\citeauthoryear{Wang et~al.,}{Wang et~al.}{2010a}]{Wang2010}
Wang J.  et~al., 2010a, Astrophys. J., 723, 620, \eprint{1008.3391},
  \doi{10.1088/0004-637X/723/1/620}

\bibitem[\protect\citeauthoryear{Wang et~al.,}{Wang et~al.}{2013b}]{Wang2013}
Wang J.  et~al., 2013b, Astrophys. J., 763, \eprint{1211.6450},
  \doi{10.1088/0004-637X/763/2/90}

\bibitem[\protect\citeauthoryear{Watkinson, Majumdar \& Pritchard}{Watkinson
  et~al.}{2017a}]{Watkinson2017}
Watkinson C.~A.,  Majumdar S.,    Pritchard J.~R.,  2017a, MNRAS, 472, 2436,
  \eprint{1705.06284}, \doi{10.1093/mnras/stx2130}

\bibitem[\protect\citeauthoryear{Watkinson et~al.,}{Watkinson
  et~al.}{2019b}]{Watkinson2018}
Watkinson C.~A.  et~al., 2019b, MNRAS, 482, 2653, \eprint{1808.02372},
  \doi{10.1093/mnras/sty2740}

\end{thebibliography}

\appendix
\section{GMCA recovery of power spectrum}
\label{appendix:gmcaPS}

\begin{figure*}
\begin{minipage}{176mm}
\begin{tabular}{c}
  \includegraphics[trim=0.0cm 0cm 0.0cm 0.0cm, clip=true, scale=0.44]{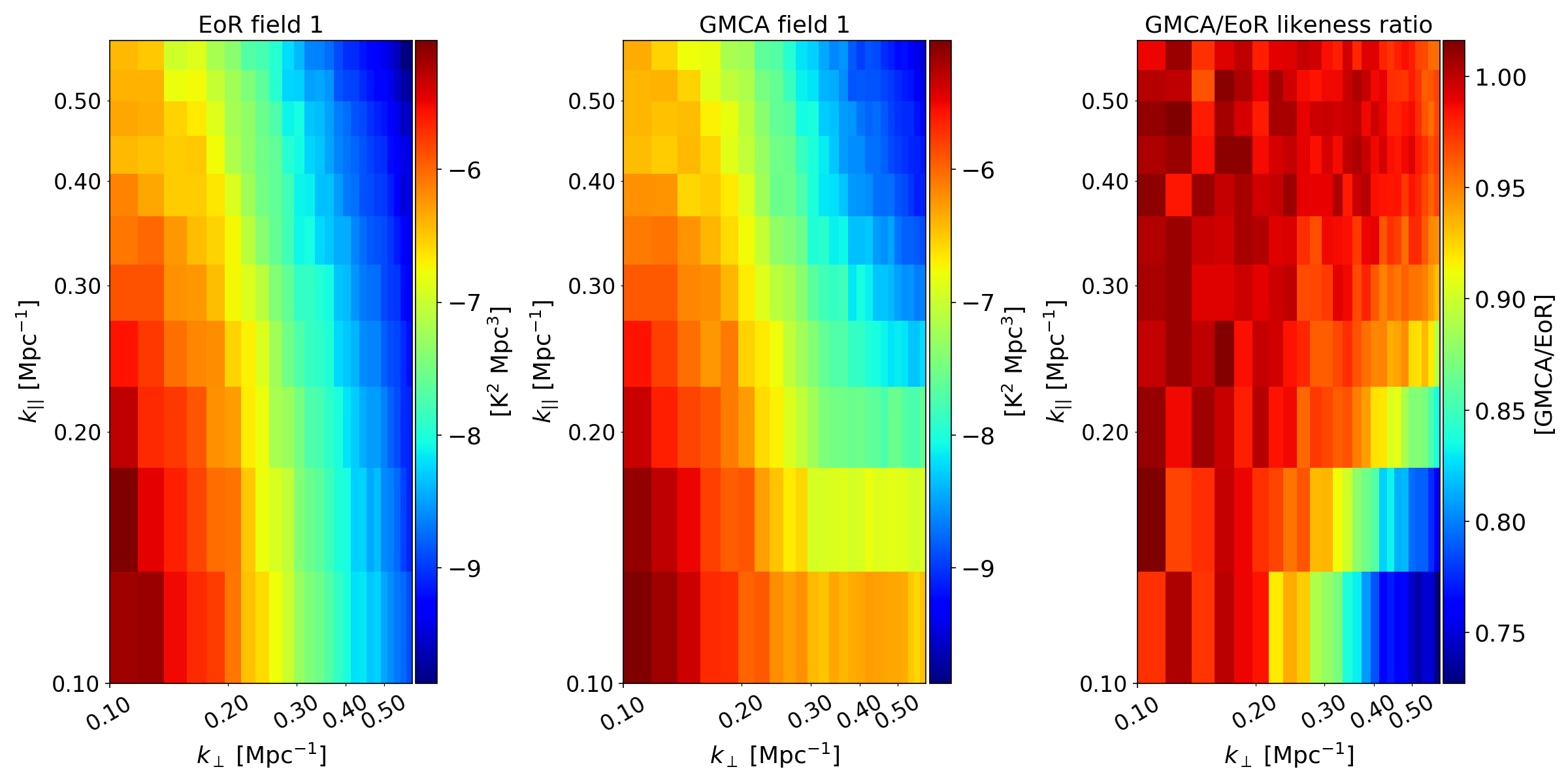}\\
\end{tabular}
\caption{Here, from left to right respectively, we compare the 2D power spectrum as measured from the "observed" EoR-only (field 1), the GMCA residuals (assuming 4 components) and the likeness ratio of the two.
We see that GMCA has recovered the observed EoR signal very well for most scales (with performance dropping within the wedge).
}\label{fig:ps_gmca_cf_obs}
\end{minipage}
\end{figure*}
In this appendix we look at how well GMCA residuals represent the EoR 2D power spectrum.
Whilst it is naturally the true signal we are interested in recovering when we perform foreground removal, it is relevant to look at how well the power spectrum of the observed EoR signal is recovered.
This is because we can in principle forward model the observation before performing parameter estimation (although in reality it is not yet clear how well and/or efficiently this can be done).
In Fig. \ref{fig:ps_gmca_cf_obs} we show the 2D power spectrum of the "observed" EoR signal (field 1) on the left, of the GMCA residuals in the middle (assuming four components to describe the foreground signal),
and the likeness ratio ($P_{\mathrm{GMCA}}/P_{\mathrm{EoR}}$) of the two
on the right.
It is clear that GMCA is able to recover the "observed" EoR signal pretty well, with the likeliness ratio being 1 for most scales, although it drops to 0.75 at scales within the wedge (which is dominated by point-source emission, see \citealt{Li2019b}).

\begin{figure*}
\begin{minipage}{176mm}
\begin{tabular}{c}
  \includegraphics[trim=0.0cm 0cm 0.0cm 0.0cm, clip=true, scale=0.44]{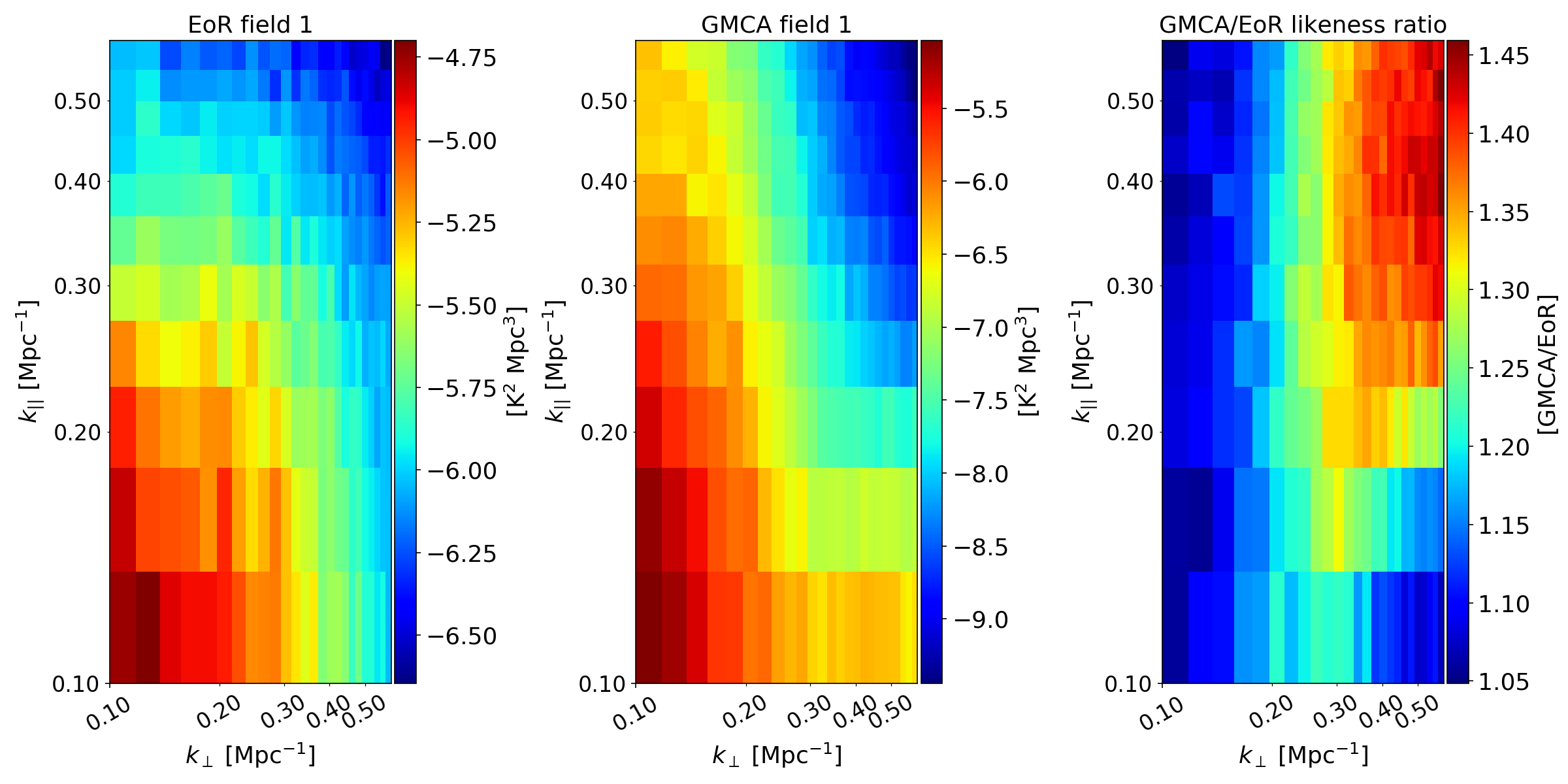}\\
\end{tabular}
\caption{Here, from left to right respectively, we compare the 2D power spectrum as measured from the clean EoR-only signal, the GMCA residuals (assuming 4 components and recovering foregrounds from "observed" field 1) and the likeness ratio of the two.
GMCA is much less effective at recovering the 2D power spectrum of the clean EoR signal.
The bispectrum is in fact better recovered for many configurations.
}\label{fig:ps_gmca_cf_clean}
\end{minipage}
\end{figure*}

GMCA residuals are not such a good representation of the clean EoR signal, as can be seen in Fig. \ref{fig:ps_gmca_cf_clean}.
This plot shows the 2D power spectrum for the clean EoR signal (left), the GMCA residuals (middle) which is as described for Fig. \ref{fig:ps_gmca_cf_clean}, and the likeness ratio (right).
The likeness ratio is greater than 1 everywhere (unlike the bispectrum for which there was a few configurations for which the likeness ratio was 1 or extremely close to). The recovery is reasonable (i.e. within 10\%) for $\kperp<0.2$, but poor otherwise.
Based on these results, the bispectrum should be just as viable a statistical probe as the power spectrum so long as we can perform foreground removal.

%%%%%%%%%%%%%%%%%%%%%%%%%%%%%%%%%%%%%%%%%%%%%%%%%%%%%%%%%%%%%%%%%%%%%%%%%%%%%%%%
%%%%%%%%%%%%%%%%%%%%%%%%%%%%%%%%%%%%%%%%%%%%%%%%%%%%%%%%%%%%%%%%%%%%%%%%%%%%%%%%

\bsp
\end{document}